\documentclass[twocolumn, twocolappendix,trackchanges]{aastex701}

\usepackage{xcolor}
\usepackage{amsmath}
\usepackage[version=4]{mhchem} 
\usepackage{rotating}
\usepackage{xcolor}

\begin{document}

\title{Irradiated Atmospheres VII. Effect of Mixing Energy Flux on Atmosphere-Interior Coupling}

\correspondingauthor{Cong Yu}
\email{yucong@mail.sysu.edu.cn}

\correspondingauthor{Dong-dong Ni}
\email{ddni@nju.edu.cn}

\author[0000-0002-0447-7207]{Wei Zhong}
\affiliation{Institute of Science and Technology for Deep Space Exploration, Nanjing University, Suzhou, 215163, PR China}
\email{}

\author[0009-0004-1986-2185]{Zhen-Tai Zhang}
\affiliation{School of Physics and Astronomy, Sun Yat-Sen University, Zhuhai, 519082, PR China}
\affiliation{CSST Science Center for the Guangdong-Hong Kong-Macau Greater Bay Area, Zhuhai, 519082, PR China}
\affiliation{State Key Laboratory of Lunar and Planetary Sciences, Macau University of Science and Technology, Macau, PR China}
\email{}

\author[0000-0002-1624-3360]{Mei Ting Mak}
\altaffiliation{Croucher Postdoctoral Fellow}
\affiliation{Atmospheric, Oceanic, and Planetary Physics Department, University of Oxford, OX1 3PU, UK}
\affiliation{Department of Physics and Astronomy, University of Exeter, Exeter EX4 4QL, UK}
\email{}

\author[0009-0009-1023-1041]{Xiang-Xing Chen}
\affiliation{Institute of Science and Technology for Deep Space Exploration, Nanjing University, Suzhou, 215163, PR China}
\email{}

\author[0000-0002-0378-2023]{Bo Ma}
 \affiliation{School of Physics and Astronomy, Sun Yat-Sen University, Zhuhai, 519082, PR China}
\affiliation{CSST Science Center for the Guangdong-Hong Kong-Macau Greater Bay Area, Zhuhai, 519082, PR China}
\email{}

\author[0000-0003-2278-6932]{Xian-Yu Tan}
\affiliation{Tsung-Dao Lee Institute \& School of Physics and Astronomy, Shanghai Jiao Tong University, Shanghai 201210, PR China}
\email{}

\author[0000-0003-0454-7890]{Cong Yu}
\affiliation{School of Physics and Astronomy, Sun Yat-Sen University, Zhuhai, 519082, PR China}
\affiliation{CSST Science Center for the Guangdong-Hong Kong-Macau Greater Bay Area, Zhuhai, 519082, PR China}
\affiliation{State Key Laboratory of Lunar and Planetary Sciences, Macau University of Science and Technology, Macau, PR China}
\affiliation{International Centre of Supernovae, Yunnan Key Laboratory, Kunming 650216, PR China}
\email{}

\author[0000-0002-0483-0445]{Dong-dong Ni}
\affiliation{Institute of Science and Technology for Deep Space Exploration, Nanjing University, Suzhou, 215163, PR China}
\affiliation{State Key Laboratory of Lunar and Planetary Sciences, Macau University of Science and Technology, Macau, PR China}
\affiliation{Shishan Laboratory, Nanjing University, Suzhou 215000, PR China}
\affiliation{Jiangsu Key Laboratory of Quantum Information Science and Technology, Nanjing University, Suzhou, 215163, PR China}
\email{}

\begin{abstract}

Highly irradiated, underdense rocky planets such as puffy Venuses may host strong coupling between their atmospheres and magma-ocean interiors. Previous studies have mainly treated this coupling through radiative transfer and chemical equilibrium, leaving the role of vertical mixing unclear. Here we quantify how energy transport associated with vertical mixing modifies the thermal structure, volatile chemistry, and emission spectra of puffy Venuses. Atmospheric circulation and gravity-wave breaking are possible dynamical sources of such transport. We show that this energy transport provides an additional greenhouse-like heating pathway by warming the deep atmosphere, thereby increasing the pressure and temperature at the magma-ocean surface. The resulting thermal changes also reshape the atmospheric composition, especially the abundances of \ce{CO} and \ce{H2O}, and change the partitioning of C-, H-, and O-bearing volatiles between the atmosphere and the magma ocean. Moreover, these changes reshape the mid-infrared planet--star flux ratio, especially at low equilibrium temperatures and in higher-mass planets. Under high C/O, it slightly increases the flux near $3.5$--$4.0\,\mu$m, but more broadly reduces atmospheric transparency and suppresses the flux. This may affect radius estimates for puffy Venuses over a limited temperature range.

\end{abstract}


\section{Introduction}

Recent JWST observations have renewed interest in highly irradiated rocky exoplanets with molten or partially molten surfaces and possible volatile atmospheres. Depending on atmospheric heat redistribution and internal heat sources, these planets may sustain either global magma oceans or dayside magma oceans, and are commonly referred to as lava worlds \citep{2025PSJ..Coy,2026arXiv260107080F}.
A subset of such planets with sufficiently extended, carbon-rich atmospheres may also experience substantial radius inflation \citep{2025PSJ..Coy,2026arXiv260107080F} and have been described as puffy Venuses \citep{2024ApJ...Peng}. Although the atmosphere is the primary observable component, the evolution of these planets is fundamentally governed by atmosphere--interior coupling.

This coupling is established at the atmosphere--magma ocean interface.
The magma ocean outgasses volatile species such as \ce{H2O}, \ce{CO}, and \ce{CO2}, while melt--gas reactions and dissolution back into the melt regulate atmospheric abundances and elemental ratios, including C/O.
In turn, the atmosphere controls volatile escape, radiative cooling, and the surface energy budget of the magma ocean.
The resulting two-way feedback shapes both atmospheric chemistry and interior evolution
\citep{Abe1986JGR....91E.291A,Elkins-Tanton2008,Schaefer2016ApJ,Nicholls2026NatAs,Nicholls2026CHILI}.
The broader role of C/O in determining terrestrial-planet composition was discussed by \citet{Bond2010ApJ}, while \citet{Madhusudhan2012ApJ758} showed that atmospheric C/O is an important diagnostic of atmospheric chemistry and can constrain planetary interiors and formation conditions.
For atmosphere--magma ocean systems specifically, studies have examined how volatile partitioning and redox evolution regulate atmospheric composition and elemental ratios
\citep{2023Natur.Young,2024ApJ...Seo,2025ApJ...Werlen,2026ApJ...999..178W}.
Although non-ideal thermodynamic effects can become important under the extreme pressures relevant to deep sub-Neptune interiors
\citep{2026ApJ...999..178W}, such effects are expected to be secondary for super-Earths and smaller rocky planets, including those considered here \citep{Bower2025ApJ}.

55~Cnc~e provides an important observational motivation for treating lava worlds as coupled interior--surface--atmosphere systems. 
JWST NIRCam and MIRI observations of its 4--12~$\mu$m thermal emission spectrum disfavor a tenuous atmosphere made only of vaporized rock. 
Instead, they support a secondary volatile atmosphere, possibly rich in \ce{CO} or \ce{CO2}, that may be supplied by magma-ocean outgassing \citep{2024Natur...Hu}. 
This interpretation shows that the atmosphere is not an isolated layer. 
Its composition can be controlled by volatile exchange with the magma ocean and the underlying interior. 
At the same time, atmospheric cooling, circulation, and escape can reshape the surface volatile budget. 
Recent models therefore emphasize the coupled evolution of dayside magma oceans, mantle convection, volatile transport, and atmospheric outgassing \citep[e.g.,][]{2026MNRAS...Meier}. 
Atmospheric spectra may thus provide a direct probe of interior--surface--atmosphere coupling and help constrain volatile inventories, C/O ratio, \ce{H2O} abundance, and molecular diagnostics such as \ce{SiH4}/\ce{CH4} 
\citep{2024ApJ...Seo,2026MNRAS...Hakim,2025ApJ...Werlen}.

To explore this coupled picture across different planetary regimes, we consider 55~Cnc~e and TOI-561~b as two test cases.
The former serves as an observational benchmark for a  relatively massive lava world with a possible secondary volatile atmosphere.
The latter provides a complementary lower-mass example of a highly irradiated, low-density super-Earth.
Recent JWST/NIRSpec dayside emission observations of TOI-561~b disfavor a bare-rock interpretation and support the presence of a thick volatile atmosphere, highlighting TOI-561~b as an important example of a highly irradiated low-mass planet with a possible volatile reservoir \citep{Teske2025ApJ...995L..39T}.
Together, these two planets allow us to examine how interior--surface--atmosphere coupling may shape observable spectra 
under different planetary masses and volatile conditions.

Existing models of atmosphere--interior coupling on lava worlds have primarily emphasized volatile partitioning, equilibrium chemistry, radiative transfer, and radiative--convective structure \citep{2024A&A...Maurice,2025MNRAS..Nicholls,2025arXiv251116142L}. More recent studies have also begun to explore the roles of tidal heating, magma-ocean tides, and magma-ocean circulation in setting the thermal and dynamical states of these planets \citep{2025ApJ...Farhat,2026arXiv260107080F,2025JOSS...Nicholls,2026arXiv260400535L}. However, the consequences of mechanically driven energy deposition within the atmosphere for atmosphere--interior coupling remain poorly understood.

\begin{figure*}
    \centering
     \includegraphics[width=1.0\linewidth]{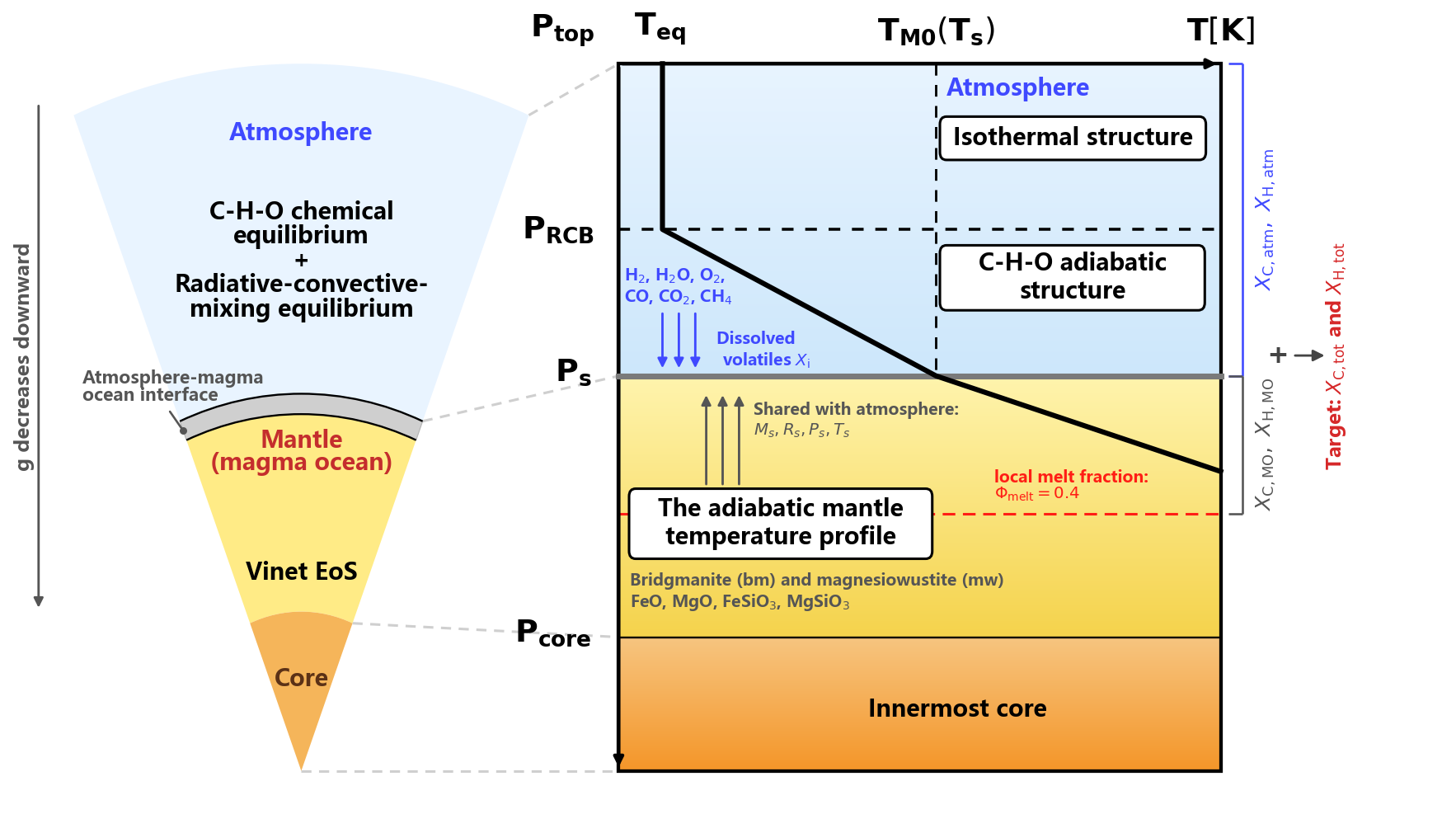}
    \caption{
    Schematic of the coupled atmosphere--mantle model.
    The light-blue region represents the atmosphere, where C--H--O chemical equilibrium and radiative--convective-mixing transport determine the composition and thermal structure.
    The yellow region represents the mantle, described by the VintEoS
    using the lower-mantle mineralogical prescription of \citet{2024ApJ...Peng},
    which is used to calculate the thermal structure and depth of the magma ocean.
    We adopt a spherically symmetric atmosphere, illustrated by the sector geometry, so that gravity is allowed to vary inward with the planetary structure rather than being held constant.
    The gray band marks the atmosphere--MO interface, where the atmosphere and  MO share the same surface pressure $P_{\rm s}$, temperature $T_{\rm s}$, radius $R_{\rm s}$, and enclosed mass $M_{\rm s}$, and where atmospheric volatiles equilibrate with dissolved volatiles in the melt.
    The red dashed curve denotes the $\phi_{\rm melt}=0.4$ contour, following \citet{2024ApJ...Peng}.
    This contour defines the base of the magma ocean.
    The region above it, where $\phi_{\rm melt}\geq0.4$, extends to the atmosphere--MO interface and has a total mass $M_{\rm MO}$.
    The schematic does not imply that the entire mantle is molten.
    The combined atmosphere--magma-ocean reservoir is used to compute the total carbon and hydrogen mass fractions, $X_{\rm C,tot}$ and $X_{\rm H,tot}$.
    The orange region is the rocky core, and the black curve indicates the assumed temperature profile.
    }
    \label{fig:structure}
\end{figure*}

One process that may be particularly important, but has not yet been explored in this context, is the deposition of mechanical energy associated with vertical mixing in stably stratified atmospheric layers. 
Such vertical mixing may arise from large-scale atmospheric circulation \citep{Zhang_2020} or from gravity-wave saturation and breaking \citep{Strobel1987Vertical,Zhang_2020}. 
In exoplanet atmospheres, circulation driven by uneven stellar heating can excite gravity waves through flow adjustment and dynamical instabilities. These waves can propagate into stably stratified layers and may eventually saturate or break as their amplitudes increase. In Earth's middle atmosphere, \citet{MedvedevKlaassen2003} showed that gravity-wave saturation and breaking produce a downward sensible-heat flux whose divergence typically heats the layers below and cools those above, while the dissipated wave kinetic energy is irreversibly converted into heat. 
Previous studies have mainly emphasized material transport and mixing associated with convection and turbulence in convective regions \citep{Zhang_2020}. 
Here, in contrast, we focus on the thermal effect of vertical energy transport in the stably stratified radiative zone and its possible role in atmosphere--interior coupling. The adopted $K_{\rm zz}^{\rm heat}$ is used to describe this thermal energy transport and is not intended to specify a unique dynamical driver. 
This energy transport may alter the thermal structure of the deep atmosphere and, in turn, modify the surface boundary condition of the magma ocean. By warming the deep atmosphere and the magma-ocean surface, this process may reduce the stellar irradiation required to sustain a given melt state or atmospheric thickness.

In this study, we investigate how gravity‐wave breaking and the associated vertical mixing influence atmosphere–interior coupling on lava worlds. We quantify the resulting modifications to magma‐ocean surface temperature, atmospheric thermal structure, and observable planetary properties, and we discuss the implications for interpreting JWST observations of hot rocky exoplanets.

The structure of this work is as follows. In \S~\ref{Sec:Method}, we describe the methodology used to simulate the coupled magma ocean--atmosphere system, including the treatment of vertical mixing. 
\S~\ref{Sec:results} presents the thermal impact of vertical mixing on the atmospheric structure, the associated changes in volatile speciation under different mantle redox conditions, and the resulting changes in the planetary emission spectrum. Finally, \S~\ref{Sec:conclusion}  provides a summary of our main findings and a discussion of their broader implications.

\section{Magma Ocean-Atmosphere model with vertical mixing }\label{Sec:Method}

We develop a coupled planetary model consisting of a carbon-bearing C--H--O atmosphere and an underlying magma ocean (MO), which are assumed to be in thermochemical equilibrium at their interface.
We focus on how energy transport associated with vertical mixing affects atmosphere--interior coupling.
The coupling framework is shown in Fig.~\ref{fig:structure}.
The model comprises three regions, of which the atmosphere and the magma ocean are modeled explicitly.
From outside to inside, these are:
(1) a radiative--convective atmosphere that includes vertical mixing and carbon-hydrogen-oxygen (C-H-O) chemical equilibrium;
(2) a magma ocean,
whose adiabatic thermal structure is calculated following \citet{2024ApJ...Peng} using the Vinet equation of state for a lower-mantle mixture of bridgmanite (bm) and magnesiowüstite (mw).
This mixture specifies the equation of state rather than the spatial extent of the MO, whose base is defined by the $\phi_{\rm melt}=0.4$ rheological boundary;
and (3) an underlying rock-like interior, which is not modeled explicitly.

This manuscript is organized as follows. \S~\ref{apdx:CE} presents the key atmospheric reactions and equilibrium relations.  \S~\ref{sec:mixing in atmosphere} describes the radiative transfer equations. \S~\ref{sec:magam_ocean_mantle} outlines the magma-ocean (lower-mantle) calculations. \S~\ref{apdx:dissolution} explains how we couple the atmosphere to the magma ocean.

\subsection{Atmospheric chemical equilibrium}\label{apdx:CE}

Our atmospheric model links six volatile species---\ce{H2}, \ce{O2}, \ce{H2O}, \ce{CO2}, \ce{CO}, and \ce{CH4}---through three gas-phase reactions \citep{Bower2022PSJ,2024ApJ...Peng}:
\begin{equation}\label{eqn:CO2}
    \ce{CO + 0.5O2} \Leftrightarrow \ce{CO2}\ ,
\end{equation}
\begin{equation}
    \ce{H2 + 0.5\,O2} \Leftrightarrow \ce{H2O}\ ,
\end{equation}
\begin{equation}
    \ce{CO2 + 2\,H2} \Leftrightarrow \ce{CH4 + O2}\ .
\end{equation}
For each prescribed C/O ratio, the corresponding C/H ratio is determined self-consistently from the coupled atmosphere--magma ocean equations.
Both elemental ratios are uniform with altitude, whereas the local equilibrium molar fractions vary with pressure and temperature as
$y_i=y_i(P,T;\mathrm{C/O},\mathrm{C/H})$.
We assume local thermochemical equilibrium throughout the atmosphere and equilibrium with the magma ocean at the atmosphere--MO interface.
This approximation is most appropriate in the hot, high-pressure deep atmosphere, where chemical equilibration is expected to be faster than vertical transport.
It may break down where chemical timescales exceed transport timescales, most plausibly in the cooler, lower-pressure upper atmosphere.
The resulting composition affects the thermodynamic properties of the atmosphere, including its density and mean heat capacity.

For an ideal, single-species gas, the specific Gibbs free energy may be written as \citep{Denbigh_1981,2016ApJ...816...Heng,2024ApJ...963..Tian}
\begin{equation}
G = G_0 + \mathcal{R}T \ln \Psi\ ,
\end{equation}
where $G_0 \equiv G(P_0)$, $P_0$ is a reference pressure (typically $1$~bar), and $\mathcal{R}$ is the universal gas constant. In this expression, $\mathcal{R}=8.314462618\,{\rm J\,mol^{-1}\,K^{-1}}$.
For an ideal gas, $\Psi = P/P_0$; the goal is to generalize $\Psi$ to gas mixtures and to non-ideal behavior. In analogy with the formalism of \citet{2024ApJ...963..Tian}, we write for species $i$ in a mixture
\begin{equation}
\Psi_i = \frac{a_i f}{f_0} = \frac{f_i}{P_0}\ ,
\end{equation}
where $f$ is the fugacity of the corresponding pure gas, $a_i$ is the activity (generalizing the mixing ratio), and $f_i \equiv a_i f$ is the fugacity of species $i$ in the mixture. We take $f_0 = P_0$ (appropriate when $P_0=1$ bar and the gas is close to ideal at the reference state). Note that $\Psi_i$ is not the activity; the two are sometimes conflated only because different factors can be grouped inside the logarithm \citep{Denbigh_1981,2016ApJ...816...Heng,2024ApJ...963..Tian}.

As an example, consider reaction~(\ref{eqn:CO2}). The Gibbs free energies of the relevant species may be written as
\begin{equation}
\begin{aligned}
G_{\ce{CO2}} & = G_{\ce{CO2},0} + \mathcal{R}T \ln \Psi_{\ce{CO2}}\ , \\
G_{\ce{O2}}  & = G_{\ce{O2},0}  + \mathcal{R}T \ln \Psi_{\ce{O2}}\ , \\
G_{\ce{CO}}   & = G_{\ce{CO},0}   + \mathcal{R}T \ln \Psi_{\ce{CO}}\ .
\end{aligned}
\end{equation}
The Gibbs free energy change for the reaction is
\begin{equation}
\begin{aligned}
\Delta G   & = G_{\ce{CO2}} - G_{\ce{CO}} - 0.5\,G_{\ce{O2}}\ , \\
\Delta G_0 & = G_{\ce{CO2},0} - G_{\ce{CO},0} - 0.5\,G_{\ce{O2},0}\ .
\end{aligned}
\end{equation}
Combining the expressions yields
\begin{equation}
\Delta G - \Delta G_0
= \mathcal{R}T \ln\!\left(\frac{\Psi_{\ce{CO2}}}{\Psi_{\ce{CO}}\,\Psi_{\ce{O2}}^{1/2}}\right)\ .
\end{equation}
At chemical equilibrium, $\Delta G = 0$, and $\Delta G_0$ is the standard Gibbs free energy change at $P_0$ (i.e., the Gibbs free energy of reaction evaluated at the reference pressure). The equilibrium constant is then
\begin{equation}
K_{\mathrm{eq}}(T) = \exp\!\left(-\frac{\Delta G_0}{\mathcal{R}T}\right)
= \frac{\Psi_{\ce{CO2}}}{\Psi_{\ce{CO}}\,\Psi_{\ce{O2}}^{1/2}}\ .
\end{equation}
Using $\Psi_i = f_i/P_0$ (with $P_0=1$ bar), we obtain
\begin{equation}\label{eqn:K}
K_{\rm eq,1}(T) = \frac{f_{\ce{CO2}}}{f_{\ce{CO}}\,f_{\ce{O2}}^{\,\,1/2}}\ .
\end{equation}
By the same procedure, the equilibrium constants for the other two reactions are
\begin{equation}
K_{\rm eq,2}(T) = \frac{f_{\ce{H2O}}}{f_{\ce{H2}}\,f_{\ce{O2}}^{\,\,1/2}}\ ,
\end{equation}
and
\begin{equation}
K_{\rm eq,3}(T) = \frac{f_{\ce{CH4}}\,f_{\ce{O2}}}{f_{\ce{CO2}}\,f_{\ce{H2}}^{\,\,2}}\ .
\end{equation}
To compute $K_{\rm eq}(T)$, we use thermochemical data from the NIST Chemistry WebBook \citep{NIST}.
For the six species considered here, the NIST Shomate-polynomial fits provide a common temperature range of $500$--$6000~\mathrm{K}$.
All atmospheric layers in our models have $T\geq500~\mathrm{K}$, so no low-temperature extrapolation is required.
For $T>6000~\mathrm{K}$, we linearly extrapolate the NIST-based thermochemical data following \citet{2024ApJ...Peng} and use the extrapolated values to calculate $K_{\rm eq}(T)$.
The resulting $K_{\rm eq}(T)$ functions for the three gas-phase reactions are shown in Fig.~\ref{fig:Keq} in the Appendix.

With \(K_{\rm eq}(T)\) specified, chemical equilibrium at each atmospheric layer can be enforced through the fugacity form of the mass-action relations. This requires expressing the fugacity of each species in terms of the local thermodynamic state \((P,T)\) and the layer composition. For a non-ideal gas, the fugacity is written as \(f=\phi P\), where \(\phi\) is the fugacity coefficient obtained from an equation of state (EoS). Under our assumptions, we approximate the fugacity of species \(i\) in the mixture as
\begin{equation}
f_i = \phi_{i,\mathrm{pure}}\, y_i\, P,
\end{equation}
where \(y_i\) is the local molar fraction and \(P\) is the total pressure at that radial location in the atmosphere. Here \(\phi_{i,\mathrm{pure}}\) is the fugacity coefficient of the corresponding pure gas evaluated at the local \((P,T)\).

We compute \(\phi_{i,\mathrm{pure}}\) using non-ideal EoS fits.
Specifically, we adopt the modified Lee--Kesler EoS developed by \citet{Duan_1992_EOS} for \ce{CO2} and \ce{H2O}, and its generalized extension for the remaining four species \citep{Duan_1996_EOS}.
The EoS of \citet{Duan_1992_EOS} is calibrated up to \(1273~\mathrm{K}\) and \(8000~\mathrm{bar}\), whereas that of \citet{Duan_1996_EOS} is calibrated up to \(2000~\mathrm{K}\) and \(25{,}000~\mathrm{bar}\).
For conditions outside these ranges, including the \(T>6000~\mathrm{K}\) conditions reached in the deepest atmospheric layers, we follow \citet{2024ApJ...Peng} and continue to evaluate the same EoS fits without modification.
Their quantitative accuracy outside the calibrated ranges is less well constrained.

The prescribed C/O ratio and the self-consistently determined C/H ratio are both assumed to be vertically uniform throughout the atmosphere.
In terms of the local equilibrium molar fractions, these constraints imply
\begin{equation}
\mathrm{C}/\mathrm{O} =
\frac{y_{\ce{CO2}}+y_{\ce{CO}}+y_{\ce{CH4}}}
{2y_{\ce{CO2}}+y_{\ce{CO}}+2y_{\ce{O2}}+y_{\ce{H2O}}}\ ,
\end{equation}
and
\begin{equation}\label{eqn:C/H}
\mathrm{C}/\mathrm{H} =
\frac{y_{\ce{CO2}}+y_{\ce{CO}}+y_{\ce{CH4}}}
{4y_{\ce{CH4}}+2y_{\ce{H2O}}+2y_{\ce{H2}}}\ .
\end{equation}
We close the system using the mole-fraction normalization condition,
$\sum_i y_i=1$.
The mass-action relations and elemental-ratio constraints in
Eqs.~(\ref{eqn:K})--(\ref{eqn:C/H}), together with this normalization condition, define a nonlinear system that we solve numerically in each atmospheric layer to obtain the local chemical-equilibrium composition,
$y_i(P,T;\mathrm{C/O},\mathrm{C/H})$.
Vertical mixing therefore affects the equilibrium abundances only indirectly through changes in the atmospheric temperature structure.

In our model, C/O is prescribed, whereas C/H is determined self-consistently from the coupled atmosphere--MO equilibrium and elemental mass balance. In particular, C/O serves as a convenient proxy for the redox state of the outgassed mixture. Because the mantle contains vastly more redox capacity than the atmosphere, it is expected to buffer and control the atmospheric redox state. In the literature, mantle redox conditions are commonly parameterized using mineral redox buffers, most often the iron--w\"ustite (IW) buffer (e.g., \citealt{Gaillard2022,Bower2022PSJ,2024ApJ...963..Tian}). However, these buffers are experimentally calibrated for terrestrial mantle conditions and temperatures below \(\sim 3000~\mathrm{K}\) (e.g., \citealt{Ballhaus1991CoMP,Hirschmann2021GeCoA}). Our deep atmospheres can exceed this range and may reach \(>6000~\mathrm{K}\) for the most strongly irradiated planets. We therefore treat C/O as an input parameter in our atmospheric calculations, rather than extrapolating mineral-buffer calibrations far beyond their validated regime.

\subsection{Vertical Mixing in Radiatively Heated Atmospheres} \label{sec:mixing in atmosphere}

In this work, we consider vertical mixing in the upper radiative zone that may be associated with atmospheric circulation or gravity-wave breaking.
We parameterize the associated turbulent heat transport using the mixing-length prescription for forced turbulence in a stably stratified atmosphere developed by \citet{Youdin2010}.
We take downward energy flux to be negative, i.e.,
\begin{equation}\label{eq:Fmix}
    F_{\rm mix}
    =-K_{\mathrm{zz}}^{\rm heat}\,\rho\,g
    \left(1-\frac{\nabla}{\nabla_{\rm ad}}\right),
\end{equation}
where $K_{\mathrm{zz}}^{\rm heat}$ is the effective eddy diffusivity associated with energy transport by vertical mixing.
In our model, $K_{\mathrm{zz}}^{\rm heat}$ is used only for energy transport and is not applied to the vertical diffusion of chemical species.
The quantities $\rho$ and $g$ are the gas density and gravitational acceleration, respectively.
The local logarithmic temperature gradient is defined as
\begin{equation}
    \nabla\equiv\frac{d\ln T}{d\ln P}.
\end{equation}
The quantity $\nabla_{\rm ad}$ is the adiabatic logarithmic temperature gradient.
The radiative gradient used to assess convective stability is evaluated locally from the radiative $P$--$T$ profile as $\nabla_{\rm rad}\equiv(d\ln T/d\ln P)_{\rm rad}$.
The gradient $\nabla$ in Equation~(\ref{eq:Fmix}) is evaluated from the current atmospheric structure.
In radiatively stable layers, $\nabla<\nabla_{\rm ad}$, so $F_{\rm mix}<0$ and the heat flux is directed downward.
Its magnitude increases with $K_{\mathrm{zz}}$ and with $1-\nabla/\nabla_{\rm ad}$, and it vanishes as $\nabla\rightarrow\nabla_{\rm ad}$.

We compute the non-ideal adiabatic gradient consistently with the real-gas EoS as
\begin{equation}
\nabla_{\rm ad}=\left.\frac{d\ln T}{d\ln P}\right|_{\rm ad}
=\left.\frac{P}{\overline{C_p}}\frac{d\overline{V}}{dT}\right|_{P},
\end{equation}
where $\overline{C_p}$ is the \emph{mean molar} heat capacity at constant pressure, and $\overline{V}$ is the \emph{mixture molar volume}. We evaluate
\begin{equation}
\overline{C_p}(T)=\sum_i y_i\,C_{p,i}(T).
\end{equation}

For each species, we compute $C_{p,i}(T)$ using a Shomate fit to tabulated NIST data \citep{NIST},
\begin{equation}
C_{p,i}(T)=A+B t+C t^2+D t^3+E t^{-2},
\end{equation}
where $t=T/1000$ and $A$--$E$ are species-dependent coefficients. We linearly extrapolate $C_{p,i}(T)$ above $6000\,\mathrm{K}$.

For each species, the temperature derivative of the molar volume at fixed pressure is
\begin{equation}
\left.\frac{dV_i}{dT}\right|_{P}
=\frac{\mathcal{R}}{P}\left[ Z_i + T\left.\frac{dZ_i}{dT}\right|_{P}\right],
\end{equation}
where $Z_i(P,T)$ is the compressibility factor of species $i$ defined by
\begin{equation}
Z_i \equiv \frac{P V_i}{\mathcal{R} T}.
\end{equation}
We precomputed tables of $Z_i(P,T)$ and $\left.(dZ_i/dT)\right|_P$ from the EoS and interpolate them during the atmospheric integration. The mixture-averaged derivative is
\begin{equation}
\left.\frac{d\overline{V}}{dT}\right|_{P}
=\left.\sum_i y_i\frac{dV_i}{dT}\right|_{P}
+\left.\sum_i V_i \frac{dy_i}{dT}\right|_{P}.
\end{equation}
The second term accounts for changes in the equilibrium composition with temperature.

We compute the non-ideal density using a real-gas equation of state (EoS),
\begin{equation}
\rho=\frac{P\,\overline{\mu}}{Z_{\rm mix}\,\mathcal{R}\,T},
\end{equation}
where $P$ is pressure, $T$ is temperature, and $\overline{\mu}=\sum_i y_i\mu_i$ is the mean molecular mass of the mixture (with molar fraction $y_i$ and molecular mass $\mu_i$ for species $i$). The mixture compressibility factor is denoted $Z_{\rm mix}(P,T)$. In practice, we obtain $Z_{\rm mix}$ from the EoS tables; when using species-based tables, we evaluate it with the same mixing rule adopted for the EoS (e.g., $Z_{\rm mix}=\sum_i y_i Z_i$ if that approximation is used consistently). We take the compressibility factors from \citet{Duan_1992_EOS,Duan_1996_EOS}.

In this section, we incorporate energy transport by vertical mixing into the atmospheric energy equilibrium.
This extends the conventional radiative--convective equilibrium (RCE) framework
\citep{Hubeny2014book,Hubeny2017MNRAS,Gandhi2017MNRAS}
to radiative--convective--mixing equilibrium (RCME)
\citep{2025A...Zhong,2025A...Zhang}.
The steady-state energy balance is
\begin{equation}
\int_0^{\infty}\kappa_\nu\left(J_\nu-B_\nu\right)\,d\nu
+\frac{g}{4\pi}\frac{dF_{\rm mix}}{dP}
+\frac{g}{4\pi}\frac{dF_{\rm conv}}{dP}
=0\, ,
\label{eq25}
\end{equation}
where $\kappa_\nu$ is the opacity, $J_\nu$ is the mean intensity, and $B_\nu$ is the Planck function.
The quantities $F_{\rm mix}$ and $F_{\rm conv}$ denote the vertical-mixing and convective energy fluxes, respectively.
In our model, $F_{\rm mix}$ is included in radiatively stable layers, whereas $F_{\rm conv}$ is included in convectively unstable layers.
Equation~(\ref{eq25}) therefore applies throughout both regions of the atmosphere.

In convectively unstable layers, we calculate $F_{\rm conv}$ using the standard  mixing-length theory (MLT)  formulation of \citet{Kippenhahn1990}, following its implementation in atmospheric models by \citet{Hubeny2014book}, \citet{Hubeny2017MNRAS}, and \citet{Gandhi2017MNRAS}.
\begin{equation}\label{eq:Fconv}
F_{\rm conv}
=
\left(\frac{gQH_{\rm P}}{32}\right)^{1/2}
\rho c_{\rm P}T
\left(\nabla-\nabla_{\rm el}\right)^{3/2}
\left(\frac{\ell}{H_{\rm P}}\right)^2 .
\end{equation}
Here, $H_{\rm P}=P/(\rho g)$ is the pressure scale height,
$Q=-(\partial\ln\rho/\partial\ln T)_{\rm P}$,
$c_{\rm P}$ is the mass-specific heat capacity at constant pressure,
$\ell$ is the mixing length, and
$\nabla_{\rm el}$ is the logarithmic temperature gradient within a convective element.
We adopt $\ell=H_{\rm P}$ following \citet{Gandhi2017MNRAS}.
The convective-element gradient and transport efficiency are determined using the corresponding standard MLT closure relations.

The mean intensity $J_\nu$ is the angle average of the specific intensity $I_\nu$ \citep{Gandhi2017MNRAS},
\begin{equation}
J_\nu=\frac{1}{2}\int_{-1}^{1} I_\nu(\mu)\, d\mu \, .
\end{equation}
Here $\mu\equiv \cos\theta$, where $\theta$ is the angle from the vertical. In a 1D, plane-parallel, steady atmosphere, $I_\nu$ satisfies the radiative transfer equation,
\begin{equation}\label{eq:RTF}
\mu\,\frac{\partial I_\nu}{\partial \tau_\nu}= I_\nu - S_\nu \, ,
\end{equation}
where $\tau_\nu$ is the optical depth and $S_\nu$ is the source function, defined as the ratio of total emissivity to total opacity.

We solve Equation~(\ref{eq25}) together with the radiative-transfer equation~(\ref{eq:RTF}) as a coupled steady-state problem. 
No time evolution or time-stepping of the radiative fluxes is involved.
A converged radiative-equilibrium atmosphere obtained with $F_{\rm mix}=F_{\rm conv}=0$ is used as the initial model.
Each iteration consists of a formal solution followed by a complete linearization step.
During the formal solution, the radiative-transfer equation is solved for the current temperature profile to obtain the corresponding radiation field.
The local temperature gradient is then recalculated from the resulting $P$--$T$ profile and compared with $\nabla_{\rm ad}$ according to the Schwarzschild criterion.
$F_{\rm mix}$ is evaluated in radiatively stable layers, while $F_{\rm conv}$ is calculated using MLT in convectively unstable layers.
Equations~(\ref{eq25}) and (\ref{eq:RTF}) are then jointly linearized, and the Rybicki reorganization scheme is used to update the radiation field and temperature structure
\citep{Hubeny2014book,Hubeny2017MNRAS,Gandhi2017MNRAS,2025A...Zhong,2025A...Zhang}.
The updated structure is used in the next iteration, and the calculation continues until
$\max|\delta T/T|<10^{-6}$.

Our radiative-transfer calculation adopts the plane-parallel approximation, while the hydrostatic structure is treated in spherical geometry following \citet{2024ApJ...Peng}.
This allows the radial coordinate, enclosed mass, and local gravitational acceleration to vary with depth and consistently connects the atmosphere to the underlying magma-ocean mantle.

Under spherical symmetry, hydrostatic equilibrium and mass conservation give
\begin{equation}
dP = -\rho(r)\,g(r)\,dr,
\end{equation}
and we relate pressure to the column-mass coordinate $m$ through
\begin{equation}
dP = g(r)\,dm,
\end{equation}
which yields
\begin{equation}
dr = -\frac{1}{\rho(r)}\,dm .
\end{equation}
The enclosed mass profile follows from mass conservation,
\begin{equation}
dM = 4\pi r^2 \rho(r)\,dr .
\end{equation}
Here the local gravitational acceleration is
\begin{equation}
g(r)=\frac{G\,M(r)}{r^2},
\end{equation}
where $G$ is the gravitational constant and $M(r)$ is the mass enclosed within radius $r$. By integrating these equations with $\rho(r)$ from the equation of state, we obtain $r(m)$ and $M(r)$.

By integrating these equations together with $\rho(r)$ from the EoS, we obtain the radial profiles $r(m)$, $M(r)$, and $g(r)$ and thereby construct the spherically symmetric atmospheric structure used to
couple the atmosphere consistently to the magma-ocean mantle.

\subsection{Magma-ocean Mantle}\label{sec:magam_ocean_mantle}
 
As illustrated in Fig.~\ref{fig:structure}, we model the planet’s  mantle as a two-phase assemblage of bridgmanite (bm) and magnesiow\"ustite (mw) \citep{Plotnykov2020MNRAS}. We compute its radial structure by integrating a set of coupled ordinary differential equations in the enclosed mass coordinate \(m\), solving simultaneously for the radius \(r\), density \(\rho\), pressure \(P\), temperature \(T\), and gravity \(g\):
\begin{equation}
    \frac{dr}{dm} = \frac{1}{4 \pi r^2 \rho}\ ,
\end{equation}
\begin{equation}
    \frac{d \rho}{dm} = - \frac{g}{4\pi r^2 \phi}\ ,
\end{equation}
\begin{equation}
    \frac{dg}{dm} = \frac{G}{r^2} - \frac{Gm}{2 \pi r^5 \rho}\ ,
\end{equation}
\begin{equation}
    \frac{dP}{dm} = - \frac{g}{4\pi r^2}\ ,
\end{equation}
\begin{equation}
    \frac{dT}{dm} = -\frac{g T \gamma}{4\pi r^2 K_{\rm s}}\ .
\end{equation}
Here \(\phi \equiv K_{\rm s}/\rho\) is the seismic parameter, where \(K_{\rm s}\) is the adiabatic bulk modulus obtained from an equation of state (EoS). We use the Vinet EoS (VinetEoS) as implemented by \citet{Valencia2007ApJ}. We also follow \citet{Plotnykov2020MNRAS} for the treatment of phase properties and thermodynamic quantities, including the thermal expansivity and heat capacities. The Gr\"uneisen parameter \(\gamma\) is computed using the prescription of \citet{Valencia2007ApJ}.

In principle, one may interpolate a mass--radius relation from the super-Earth model grid of \citet{Plotnykov2020MNRAS} to obtain \(R_{\rm s}\) for a given rocky-planet mass \(M_{\rm s}\).
We neither use the interior grid of \citet{Plotnykov2020MNRAS} nor adopt the Earth-like core mass fraction (CMF) of $32.5\%$ assumed in that grid.
For simplicity, we instead adopt the empirical mass--radius scaling reported by \citet{2024ApJ...Peng} and verified there to have only a minor impact on the results:
\begin{equation}\label{eq:Ms_Rs}
\frac{M_{\rm s}}{M_{\oplus}}
=
\left(\frac{R_{\rm s}}{R_{\oplus}}\right)^{1/0.27}.
\end{equation}
For a given $M_{\rm s}$, Equation~(\ref{eq:Ms_Rs}) provides an estimate of $R_{\rm s}$ at the mantle--atmosphere interface.
In our calculation, this estimated radius serves only as the outer-boundary radius used to initialize the subsequent iterative calculation of the mantle thermodynamic profiles.
The choice of CMF does not affect the results of our mantle calculations.
The boundary pressure and temperature at this interface are provided by the atmospheric model.
The corresponding boundary density is obtained by inverting the Vinet EoS at \((P,T)\).
For the first iteration, we initialize the boundary gravity as \(g_{\rm s}=GM_{\rm s}/R_{\rm s}^{2}\), and then integrate the structure equations inward.

We quantify mantle melting using solidus and liquidus curves,
$T_{\rm sol}(P)$ and $T_{\rm liq}(P)$.
We adopt peridotite fits at low pressure
\citep{MORSCHHAUSER_2011_UM_Melting}
and chondritic fits up to $140~\mathrm{GPa}$
\citep{Montreux_16_deep_MO_melting}, which we extend linearly to higher pressures.
The local melt fraction is
\begin{equation}
\phi_{\rm melt}
=
\frac{T-T_{\rm sol}}{T_{\rm liq}-T_{\rm sol}}.
\end{equation}
Here, $\phi_{\rm melt}$ denotes the local melt fraction and should not be confused with the mantle-averaged melt fraction $\Phi$.
Following \citet{2024ApJ...Peng}, we use $\phi_{\rm melt}=0.4$ only to locate the rheological base of the MO.
We calculate $M_{\rm MO}$ as the total mantle mass of the overlying region, where $\phi_{\rm melt}\geq0.4$.
Individual layers are not weighted by their local melt fractions.
Thus, $M_{\rm MO}$ represents an effective, well-mixed, rheologically mobile reservoir rather than the mass of liquid melt alone.
We assume that the volatile abundances determined at the atmosphere--MO interface apply throughout this reservoir.
This approximation may overestimate the volatile storage assigned to the effective MO relative to a melt-fraction-weighted treatment.

The model does not explicitly treat solid--melt volatile partitioning or retention in trapped interstitial melt or melt inclusions.
These processes could sequester volatiles in the deeper mantle, reducing the inventory available for atmosphere--MO equilibration and potentially decreasing the atmospheric volatile mass and surface pressure \citep{Nicholls2026CHILI}.
The net effect of these omitted processes and the effective-reservoir approximation is not quantified here.

\subsection{MO--atmosphere C and H mass balance}\label{apdx:dissolution}

We model chemical exchange between the atmosphere and the magma ocean (MO) by assuming rapid mixing at their interface, so that gas and melt are in equilibrium at the MO surface, as in previous magma-ocean studies (e.g., \citealt{Bower2022PSJ,Lichtenberg_21_Solubility,Gaillard_22_Redox_MO_degassing}). Under this assumption, the partitioning of C- and H-bearing volatiles between dissolved and outgassed reservoirs is set by surface equilibrium together with solubility in the melt.

We parameterize solubility using the Henry-law fits of \citet{Lichtenberg_21_Solubility}. For each species \(i\),
\begin{equation}\label{eqn:solubility}
X_i = a\,(y_{i,s} P_s)^{b}\ ,
\end{equation}
where \(X_i\) is the dissolved mass fraction (by weight) in the MO, \(a\) and \(b\) are empirical constants, and \(P_s\) and \(y_{i,s}\) are the total pressure and molar fraction at the MO surface. The values of \(P_s\) and \(y_{i,s}\) are taken from the atmospheric solution.

The masses of dissolved elemental carbon and hydrogen in the MO are
\begin{equation}
M_{C,{\rm MO}} = M_{\rm MO}\sum_i k_{i,C}\,\frac{\mu_C}{\mu_i}\,X_i\ ,
\end{equation}
\begin{equation}\label{eqn:MO_volatiles}
M_{H,{\rm MO}} = M_{\rm MO}\sum_i k_{i,H}\,\frac{\mu_H}{\mu_i}\,X_i\ ,
\end{equation}
where \(k_{i,C}\) and \(k_{i,H}\) are the numbers of C and H atoms in species \(i\), and \(\mu_C\), \(\mu_H\), and \(\mu_i\) are molar masses.

We then impose conservation of the total C and H inventories of the coupled MO--atmosphere system,
\begin{equation}\label{eqn:x_tot}
X_{C,{\rm tot}} = \frac{M_{C,{\rm MO}} + M_{C,{\rm atm}}}{M_{\rm tot}}\ ,
\end{equation}
\begin{equation}\label{eqn:x_tot2}
X_{H,{\rm tot}} = \frac{M_{H,{\rm MO}} + M_{H,{\rm atm}}}{M_{\rm tot}}\ ,
\end{equation}
where \(M_{C,{\rm atm}}\) and \(M_{H,{\rm atm}}\) are the atmospheric masses of elemental carbon and hydrogen obtained from the atmospheric solution, and \(M_{\rm tot}\) is the total mass of the MO plus the atmosphere.

\subsection{MO--Atmosphere Coupled Solver}

\begin{figure}
    \centering
    \includegraphics[width=1.0\linewidth]{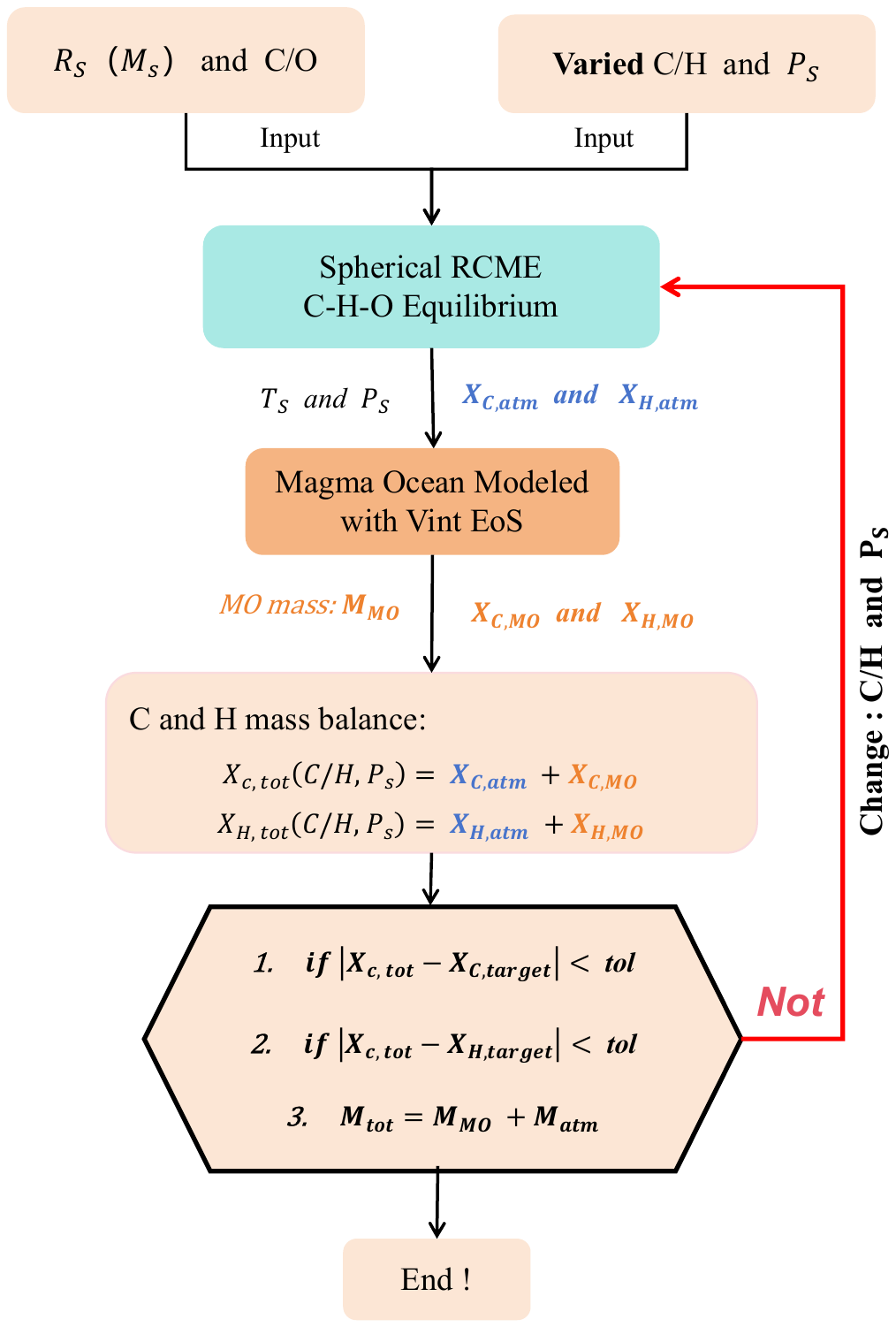}
    \caption{
    Flowchart of the MO–Atmosphere coupled solver showing the various function invocations.
    The error is defined as the norm of the residual vector between the computed total abundances, $(X_{\mathrm{C,tot}}, X_{\mathrm{H,tot}})$, and the target abundances, $(X_{\mathrm{C,target}}, X_{\mathrm{H,target}})$. 
    The optimization stops when the residual norm falls below a tolerance of $10^{-6}$.
    }
    \label{fig:flowchart}
\end{figure}

We construct a self-consistent \texttt{MO + atmosphere} model that solves the atmospheric and MO structures and partitions the prescribed bulk carbon and hydrogen inventories between the two reservoirs.
The MO provides the lower boundary of the atmosphere and stores dissolved C- and H-bearing volatiles.
The atmosphere--MO interface is defined by common values of $M_{\rm s}$, $R_{\rm s}$, $P_{\rm s}$, and $T_{\rm s}$.
At this interface, the atmospheric composition and dissolved volatile abundances are coupled through chemical and solubility equilibrium.
The overall configuration of the model is shown in Fig.~\ref{fig:structure}.
Of the six atmospheric species, \ce{CO}, \ce{CO2}, \ce{CH4}, \ce{H2}, and \ce{H2O} are allowed to dissolve in the melt.
Following \citet{2024ApJ...Peng}, we calculate their dissolved abundances using the Henry's-law solubility relations compiled by \citet{Lichtenberg2021JGRE}.
\ce{O2} is included in the atmospheric chemical-equilibrium calculation but is assumed to be insoluble.

We use the total carbon and hydrogen abundances, $X_{\rm C,tot}$ and $X_{\rm H,tot}$, to test whether a model satisfies a prescribed bulk volatile inventory. These abundances include both atmospheric and MO contributions. The atmospheric and dissolved MO masses of carbon and hydrogen are normalized by the total mass participating in volatile exchange, $M_{\rm tot}$. As reference values, bulk Earth has $(X_{\rm C,tot},X_{\rm H,tot})\simeq(500\,,200)~\mathrm{ppmw}$ \citep{MARTY_2012_CHN_Earth, Marty_2020_CN_ratio_Earth, Hirschmann_2018_deep_Earth_HCN}. Ordinary chondrites have $(1200\,,459)~\mathrm{ppmw}$, whereas enstatite chondrites have $(4100\,,1309)~\mathrm{ppmw}$ \citep{Schaefer17}. We adopt the ordinary-chondrite values as our fiducial targets.

The numerical procedure is shown in Fig.~\ref{fig:flowchart}. We first estimate the initial mass coordinate and radius of the atmosphere-MO interface using the empirical mass--radius relation for a bare rocky planet, given by Equation~(\ref{eq:Ms_Rs}). We then choose trial values of the lower-boundary pressure $P_{\rm s}$ and C/H ratio, and solve a spherically symmetric atmospheric model at fixed C/O, with C/O = 0.5 or 0.8. This module includes C--H--O chemical equilibrium and RCME,
and returns the interface state, $P_{\rm s}$ and $T_{\rm s}$. The temperature $T_{\rm s}$ is also the outer-boundary temperature of the magma ocean, $T_{\rm MO}$. In spherical geometry, gravity varies with radius and enclosed mass rather than being fixed.

For the atmospheric calculation, we set $T_{\rm int}=0~\mathrm{K}$ and use $T_{\rm eq}=1000~\mathrm{K}$ as the fiducial equilibrium temperature, with additional tests near $1200~\mathrm{K}$.
Although lower than the equilibrium temperatures of TOI-561~b and 55~Cnc~e, these values are adopted as controlled test cases rather than planet-specific inputs.
We use $T_{\rm eq}=1000~\mathrm{K}$ to facilitate comparison with the lower-temperature models of \citet{2024ApJ...Peng}, while the $1200~\mathrm{K}$ calculations provide nearby tests of the temperature dependence of vertical mixing.
The top boundary is $P_{\rm top}=0.1~\mathrm{Pa}$, $\tau_{v,{\rm top}}=10^{-9}$ and $\tau_{th,{\rm top}}=\tau_{v,{\rm top}}/\gamma$, where $\gamma=2.28\times10^{-3}$. The top temperature follows Equation~(28) of \citet{Guillot2010}, with $\mu_*=1/\sqrt{3}$ and $T_{\rm irr}^4=4T_{\rm eq}^4$. 
To incorporate vertical-mixing energy transport self-consistently, we solve the coupled radiative-transfer and energy-balance equations using the Rybicki complete-linearization scheme \citep{2025A...Zhong,2025B...Zhong,2025A...Zhang,2025B...Zhang} rather than the semi-grey ODE integration used by
\citet{2024ApJ...Peng}.
We adopt $\kappa_v=7.524\times10^{-6}$ and $\kappa_{\rm th}=3.3\times10^{-3}~\mathrm{cm^2,g^{-1}}$.

We next use $P_{\rm s}$, $T_{\rm s}$, and the interface mass coordinate and radius as the outer-boundary conditions for the MO.
The volatile-bearing mantle is computed with the VintEoS.
This step determines the internal T--P structure, the melt distribution and the amount of mantle material available for volatile storage.
We directly prescribe a core-to-bare-rocky-planet mass ratio of $0.25$, such that the mantle contains the remaining $75\%$ of the rocky-planet mass.
The mantle thermodynamic structure is then calculated self-consistently for the prescribed outer-boundary conditions.

As shown in Fig.~\ref{fig:structure}, we define the base of the MO at the local $\phi_{\rm melt}=0.4$ rheological transition.
The corresponding $M_{\rm MO}$ is the total mantle mass of the overlying region, where $\phi_{\rm melt}\geq0.4$.
Individual layers are not weighted by their local melt fractions.
Together with the atmospheric mass $M_{\rm atm}$, this gives $M_{\rm tot}$.
We then compute the C and H masses in the atmosphere and in the MO, and obtain $X_{\rm C,tot}$ and $X_{\rm H,tot}$.
The prescribed volatile inventories are therefore partitioned only between the atmosphere and the effective MO, with no separate deep-mantle reservoir.

If the calculated $X_{\rm C,tot}$ and $X_{\rm H,tot}$ do not match the target values, we update $P_{\rm s}$ and C/H and repeat the calculation. The iteration stops when the interface state is self-consistent, the atmosphere and MO are in volatile equilibrium, and the total C and H abundances match the ordinary-chondrite targets.

\section{Results}\label{Sec:results}

To examine the role of energy input from vertical mixing in atmosphere–interior coupling, this section is organized into three parts. \S~\ref{Sec:mixing in atomospheres} discusses the impact of vertical mixing on the planetary atmosphere. \S~\ref{Sec:mixing on the CMB} investigates how vertical mixing modulates mantle redox control on volatile speciation. \S~\ref{Sec:mixing on spectrum} analyzes the effect of vertical mixing on the emission spectrum.

\begin{figure*}
    \centering
    \includegraphics[width=1.0\linewidth]{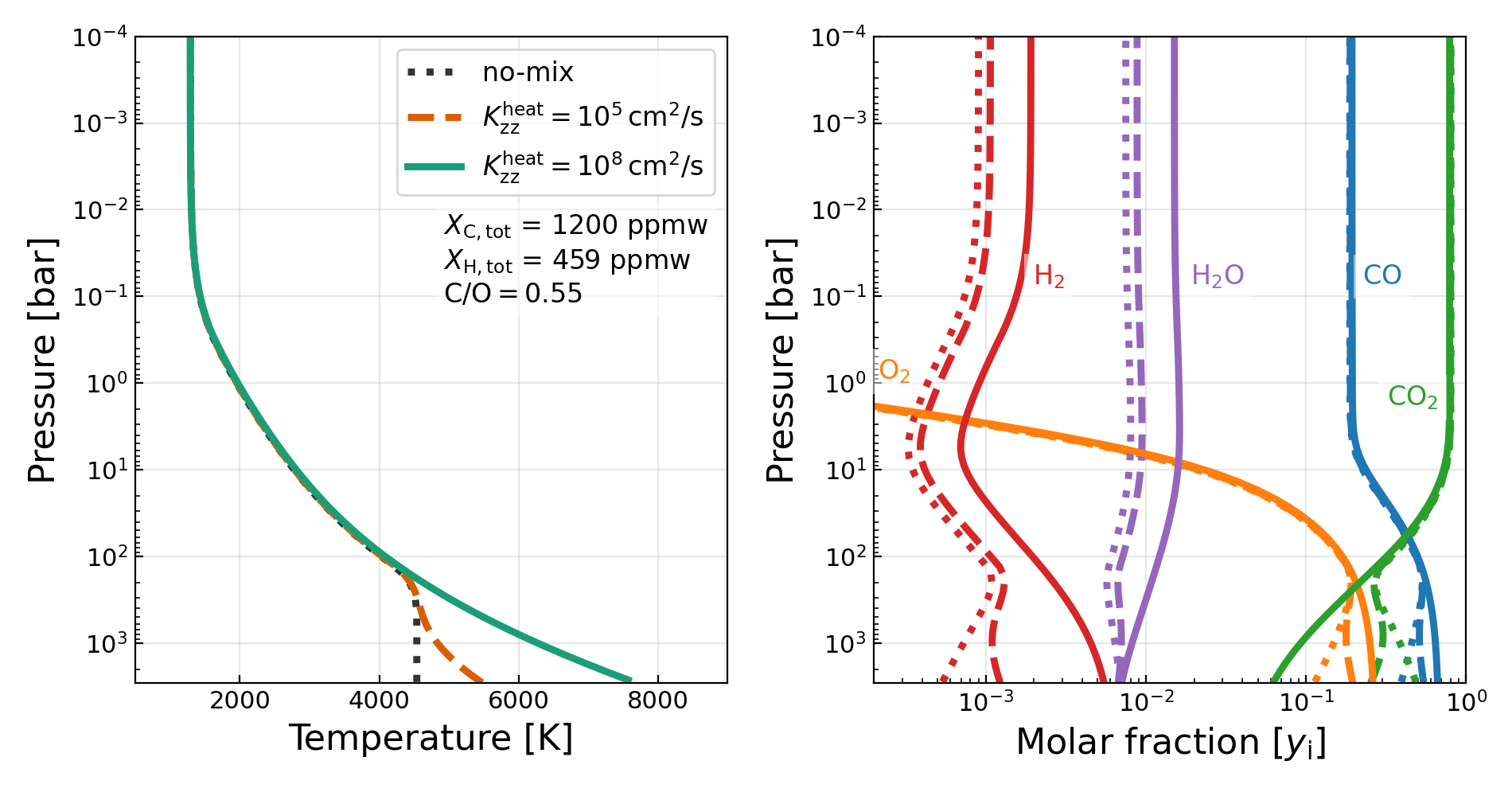}
    \caption{Vertical mixing alters the thermal structure and chemical profiles of an Earth-mass planet. Shown is a planet with solar C/O $=0.55$ and equilibrium temperature $T_{\mathrm{eq}}=1000\,\mathrm{K}$, assuming OC total C and H abundances. \textit{Left:} Pressure--temperature profiles for a no-mixing case (black dotted line) and for two vertically mixed cases with $K_{\rm zz}^{\rm heat}=10^{5}\,\mathrm{cm^{2}\,s^{-1}}$ (orange dashed line) and $K_{\rm zz}^{\rm heat}=10^{8}\,\mathrm{cm^{2}\,s^{-1}}$ (blue dashed line). \textit{Right:} Corresponding molar fraction profiles, using the same line styles to indicate each mixing scenario. Colors denote species: \ce{O2} (orange), \ce{H2} (red), \ce{H2O} (purple), \ce{CO2} (blue) and \ce{CO} (green).}
    \label{fig2:C-H-O structure}
\end{figure*}

\begin{figure*}
    \centering
    \includegraphics[width=1.0\linewidth]{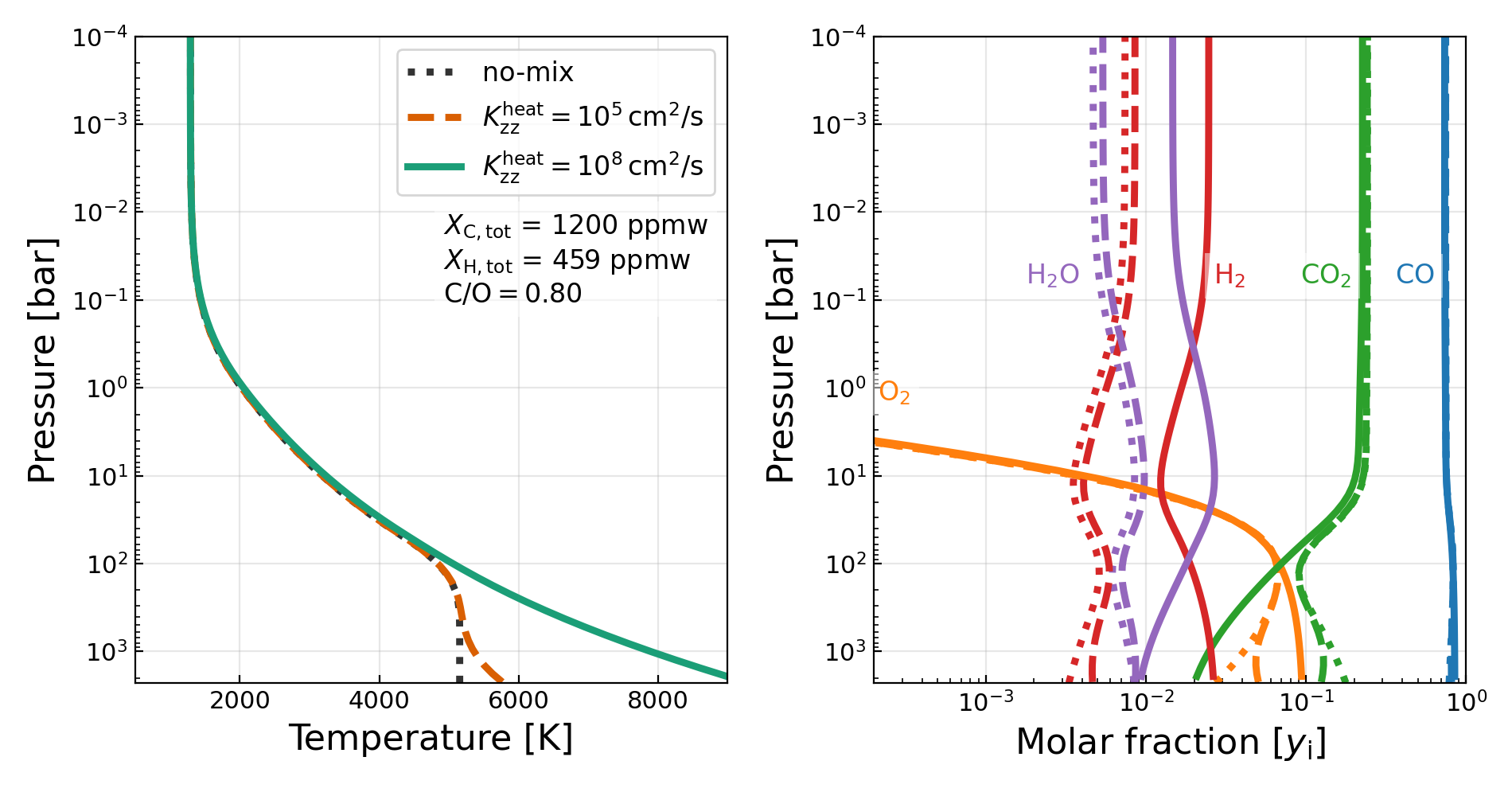}
    \caption{Same as Fig.~\ref{fig2:C-H-O structure}, but with a higher C/O ratio. Identical to Fig.~\ref{fig2:C-H-O structure}, except C/O is set to $0.80$ (line styles and colour coding are unchanged).}
    \label{fig3:C-H-O structure with C/O = 0.80}
\end{figure*}

\subsection{Atmospheric Structure with Mixing in Coupled \texttt{MO+atmosphere} Models}\label{Sec:mixing in atomospheres}

In this section, we test how vertical mixing modifies the atmospheric structure. 
The atmospheric profiles shown here are taken from the coupled \texttt{MO+atmosphere} solutions, not from isolated atmosphere models. 
The magma ocean affects the atmosphere through the shared interface state and the volatile budget. 
After C- and H-bearing species dissolve into the melt, $X_{\rm C,tot}$ and $X_{\rm H,tot}$ constrain the atmospheric C/H ratio, which controls the C--H--O chemical distribution and modifies the atmospheric density, heat capacity and $T$--$P$ structure.    
We adopt C/O$=0.55$, $T_{\rm eq}=1000~\mathrm{K}$, and bulk abundances $X_{\rm C,tot}=1200~\mathrm{ppmw}$ and $X_{\rm H,tot}=459~\mathrm{ppmw}$. The results are shown in Fig.~\ref{fig2:C-H-O structure}. In the left panel, the black dotted curve is the non-mixing case and serves as our baseline.

We begin by validating the no-mixing limit through comparison with the results reported by \citet{2024ApJ...Peng}. They state an analytic expression for the top boundary temperature,
\begin{equation}
    T_{\rm top}=\left[\frac{1}{2}T_{\rm int}^4+\frac{2}{\sqrt{3}}T_{\rm eq}^4\left(1+\frac{\sqrt{3}}{2}\gamma\right)\right]^{1/4},
\end{equation}
which is derived from the same CE radiative-equilibrium relation used here. For a given $T_{\rm eq}$, together with $T_{\rm int}$ and $\gamma$, this relation fixes $T_{\rm top}$ and thus the temperature level near the top of the atmosphere. In our comparisons, the plotted profile in \citet{2024ApJ...Peng} suggests a top-boundary convention that is not fully specified in the text. 
To ensure a like-for-like comparison, we adopt a top-boundary condition consistent with the temperature level implied by their plotted profile. With this choice, our no-mixing solution at $T_{\rm eq}=1000~\mathrm{K}$ reproduces both the  $P$--$T$ structure and the major-species mole-fraction profiles of the planet~a case shown in Fig.~1 of \citet{2024ApJ...Peng}.
This agreement confirms that our implementation matches the established benchmark in the no-mixing limit.

Vertical mixing is not generally expected to be uniform with height.
For example, \citet{Charnay2015} found an approximate scaling $K_{\rm zz}\propto P^{-0.4}$ for GJ~1214b.
Height-dependent profiles of the form $K_{\rm zz}^{\rm heat}\propto[P/(1\,{\rm bar})]^{\zeta}$ were examined in our previous study using the same vertical-mixing energy-transport framework \citep{2025A...Zhong}.
In the cases examined there, both increasing and decreasing  $K_{\rm zz}^{\rm heat}$ with altitude retain the lower-atmosphere warming found in the constant- $K_{\rm zz}^{\rm heat}$ case.
When  $K_{\rm zz}^{\rm heat}$ increases with altitude, this warming is weaker than in the constant- $K_{\rm zz}^{\rm heat}$ case.
When  $K_{\rm zz}^{\rm heat}$ decreases with altitude, the warming is stronger.
A sufficiently steep increase in  $K_{\rm zz}^{\rm heat}$ with pressure may also generate a temperature inversion in the upper atmosphere.
Thus, in those calculations, the vertical variation of  $K_{\rm zz}^{\rm heat}$ changes the magnitude and vertical distribution of the thermal response without changing the tendency to heat the lower atmosphere.
We adopt a constant  $K_{\rm zz}^{\rm heat}$ in the present study to isolate the effect of the overall mixing strength.

The values $K_{\rm zz}^{\rm heat}=10^{5}$ and $10^{8}\,\mathrm{cm^{2}\,s^{-1}}$ are adopted as representative weak- and strong-mixing cases for the sensitivity analysis rather than as observationally constrained values for a specific planet.
The temperature profiles for these two cases are shown in the left panel of Fig.~\ref{fig2:C-H-O structure}, where the black dotted curve denotes the non-mixing reference.
An additional test with $K_{\rm zz}^{\rm heat}=10^{9}\,\mathrm{cm^{2}\,s^{-1}}$ produces further warming of the atmosphere--magma-ocean interface, with no clear evidence of saturation up to this value.
At still larger  $K_{\rm zz}^{\rm heat}$, we could not obtain numerically converged solutions satisfying the imposed energy balance, so whether the interface warming eventually saturates remains undetermined.

In the no-mixing case, the atmosphere follows the layered picture in Fig.~\ref{fig:structure}. A near-isothermal region at low pressure transitions into a deeper, composition-influenced, near-adiabatic structure. At higher pressures, the profile also develops an additional radiative-controlled segment. This segment appears as a new near-isothermal region in the deep atmosphere.

Vertical mixing changes the deep thermal structure in a systematic way. 
Following \citet{Youdin2010}, we refer to this deep warming as the mechanical greenhouse effect, consistent with Papers~I--II \citep{2025A...Zhong,2025A...Zhang}.
The term reflects that vertical mixing drives a downward heat flux that partially offsets outward radiative cooling and retains heat at depth, rather than producing warming through radiative opacity.
For $K_{\rm zz}^{\rm heat}=10^{5}~\mathrm{cm^{2}\,s^{-1}}$ (orange dashed curve), the upper and middle atmosphere remain close to the no-mixing profile.
The curve begins to depart at $P\sim2\times10^{2}~\mathrm{bar}$ and the deep atmosphere becomes warmer.
For $K_{\rm zz}^{\rm heat}=10^{8}~\mathrm{cm^{2}\,s^{-1}}$ (green solid curve), the departure occurs at lower pressure.
The warming at high pressure is much larger.
Overall, stronger vertical mixing  strengthens the mechanical greenhouse effect and raises deep-atmosphere temperatures well above the no-mixing reference.

The right panel of Fig.~\ref{fig2:C-H-O structure} shows how vertical mixing changes the molar fraction profiles, $y_i$, in our C--H--O equilibrium model. Line styles match the left panel. The dotted curve is the no-mixing case, the dashed curve is $K_{\rm zz}^{\rm heat}=10^{5}~\mathrm{cm^{2}\,s^{-1}}$, and the solid curve is $K_{\rm zz}^{\rm heat}=10^{8}~\mathrm{cm^{2}\,s^{-1}}$. In the no-mixing limit, the $y_i$ profiles reproduce the results of \citet{2024ApJ...Peng}, which provides a baseline check of our implementation.

We track six species in the C--H--O network, but omit \ce{CH4} in the right panel of Fig.~\ref{fig2:C-H-O structure} because its abundance is negligible.
The dotted curves show the no-mixing case, whereas the dashed and solid curves correspond to increasing values of  $K_{\rm zz}^{\rm heat}$.
In our model,  $K_{\rm zz}^{\rm heat}$ is applied only to energy transport and does not directly transport chemical species.
Within each model, C/O and C/H are uniform with height, and all species abundances are calculated in local thermochemical equilibrium.
The abundance differences between the  $K_{\rm zz}^{\rm heat}$ cases therefore arise from mixing-induced changes in the temperature structure rather than from direct compositional mixing.

For weak mixing, $K_{\rm zz}^{\rm heat}\lesssim10^{5}~\mathrm{cm^{2}\,s^{-1}}$, the upper and middle atmosphere show only minor changes in molar fraction.
The main differences appear in the deep atmosphere at $P\gtrsim2\times10^{2}~\mathrm{bar}$.
At a given pressure in this region, the warmer temperature structure in the mixing cases shifts the local equilibrium composition toward higher abundances of \ce{H2}, \ce{O2}, and \ce{CO}, with the strongest increase in \ce{H2}.
The \ce{H2O} abundance is also slightly enhanced relative to the no-mixing case over part of the deep atmosphere, but it decreases toward higher pressures along each profile, and the different  $K_{\rm zz}^{\rm heat}$ cases converge near the lower boundary.
By contrast, the equilibrium abundance of \ce{CO2} is lower in the mixing-modified thermal structure and shows the largest decrease among the carbon-bearing species shown.

Stronger mixing amplifies these trends. For $K_{\rm zz}^{\rm heat}=10^{8}~\mathrm{cm^{2}\,s^{-1}}$, \ce{H2} and \ce{H2O} increase throughout most of the column, although \ce{H2O} approaches the no-mixing level at the highest pressures. \ce{CO} and \ce{O2} still change only modestly, while \ce{CO2} is strongly depleted in the deep atmosphere.

These compositional shifts feed back on the lower boundary of the atmosphere. As the deep composition changes, the interface conditions $(P_s, T_s)$ required for a self-consistent match to the volatile-bearing mantle also change. This is expected because the coupled MO--atmosphere solution is ultimately constrained by the prescribed bulk carbon and hydrogen inventory.

Building on the C/O$=0.55$ baseline, we next test how the input C/O ratio modifies the atmosphere and its coupling to the mantle. Changing C/O shifts C--H--O equilibrium partitioning and alters the deep thermal structure, which in turn affects the interface conditions $(P_s, T_s)$. Fig.~\ref{fig3:C-H-O structure with C/O = 0.80} shows the case C/O$=0.80$. We use the same axis limits as Fig.~\ref{fig2:C-H-O structure} to enable a direct comparison.

The temperature response is strongest at high pressure in Fig.~\ref{fig3:C-H-O structure with C/O = 0.80}. Even without vertical mixing (black dotted curve), C/O $=0.80$ produces a hotter deep radiative, near-isothermal region than in Fig.~\ref{fig2:C-H-O structure}, with temperatures exceeding $5000~\mathrm{K}$. The upper atmosphere changes little, consistent with irradiation control at low pressure. When vertical mixing is included, the mechanical-greenhouse heating of the lower atmosphere becomes larger than in the C/O $=0.55$ case. The temperature profile departs from the no-mixing reference at lower pressure and reaches higher temperatures at depth.

Compared with the C/O = 0.50 models in Fig.~\ref{fig2:C-H-O structure}, the C/O = 0.80 models in Fig.~\ref{fig3:C-H-O structure with C/O = 0.80} show a similar response to vertical mixing, but with stronger changes in several species.
As  $K_{\rm zz}^{\rm heat}$ increases, the molar fraction of \ce{CO2} is more strongly depleted in the deep atmosphere, \ce{H2} is more enhanced, and \ce{CO} changes only modestly.
The behavior of \ce{H2O} is not monotonic and depends on the mixing strength.
For the strongest mixing case, $K_{\rm zz}^{\rm heat}=10^8~\mathrm{cm^2\,s^{-1}}$, \ce{H2O} decreases toward the highest-pressure region, whereas the two weaker mixing cases show the opposite trend.
\ce{O2} increases mainly at depth, consistent with high-temperature dissociation and the resulting shift in local thermochemical equilibrium among oxygen-bearing species.

Overall, higher C/O amplifies deep-atmosphere heating and strengthens the associated compositional shifts.
For example, at $K_{\rm zz}^{\rm heat}=10^8~\mathrm{cm^2\,s^{-1}}$ and $P=10^3~\mathrm{bar}$, the $\mathrm{C/O}=0.80$ model is approximately $1800~\mathrm{K}$ hotter than the $\mathrm{C/O}=0.50$ model.
This temperature contrast becomes larger toward the atmosphere--MO interface, leading to different self-consistent values of $(P_{\rm s},T_{\rm s})$ and making the coupled solution strongly sensitive to C/O.

\subsection{Vertical Mixing  Mantle Redox Controls on Volatile Speciation}\label{Sec:mixing on the CMB}

\begin{figure*}
    \centering
    \includegraphics[width=1.0\linewidth]{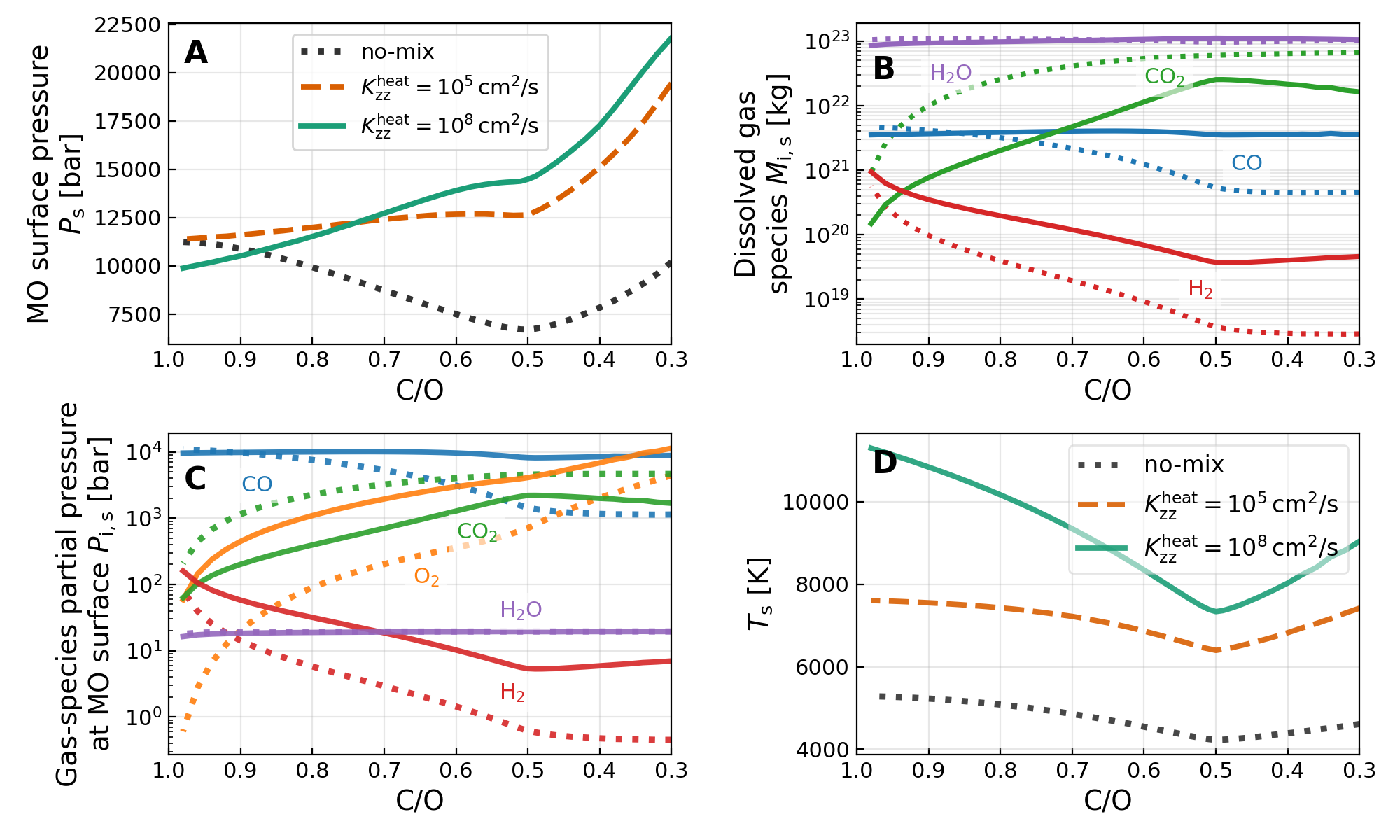}
    \caption{Impact of vertical mixing on the partitioning of volatile carbon and hydrogen as a function of the C/O ratio in a puffy, Venus-like exoplanet analogous to 55~Cnc~e, assuming OC abundances. Dotted lines indicate the case without vertical mixing, while dashed and solid lines correspond to mixing strengths of $K_{\rm zz}^{\rm heat}=10^5\,{\rm cm^2\,s^{-1}}$ and $10^8\,{\rm cm^2\,s^{-1}}$, respectively. \textbf{(A)} MO surface pressure $P_{\rm s}$ as a function of C/O ratio. \textbf{(B)} Mass of dissolved gas species $M_{\rm i,s}$ at the MO surface. \textbf{(C)} Partial pressures of major gas species $P_{\rm i, s}$ at MO surface.
    \textbf{(D)} MO surface temperature $T_{\rm s}$ as a function of C/O ratio.}
    \label{figure4}
\end{figure*}

\begin{figure}
    \centering
    \includegraphics[width=1.0\linewidth]{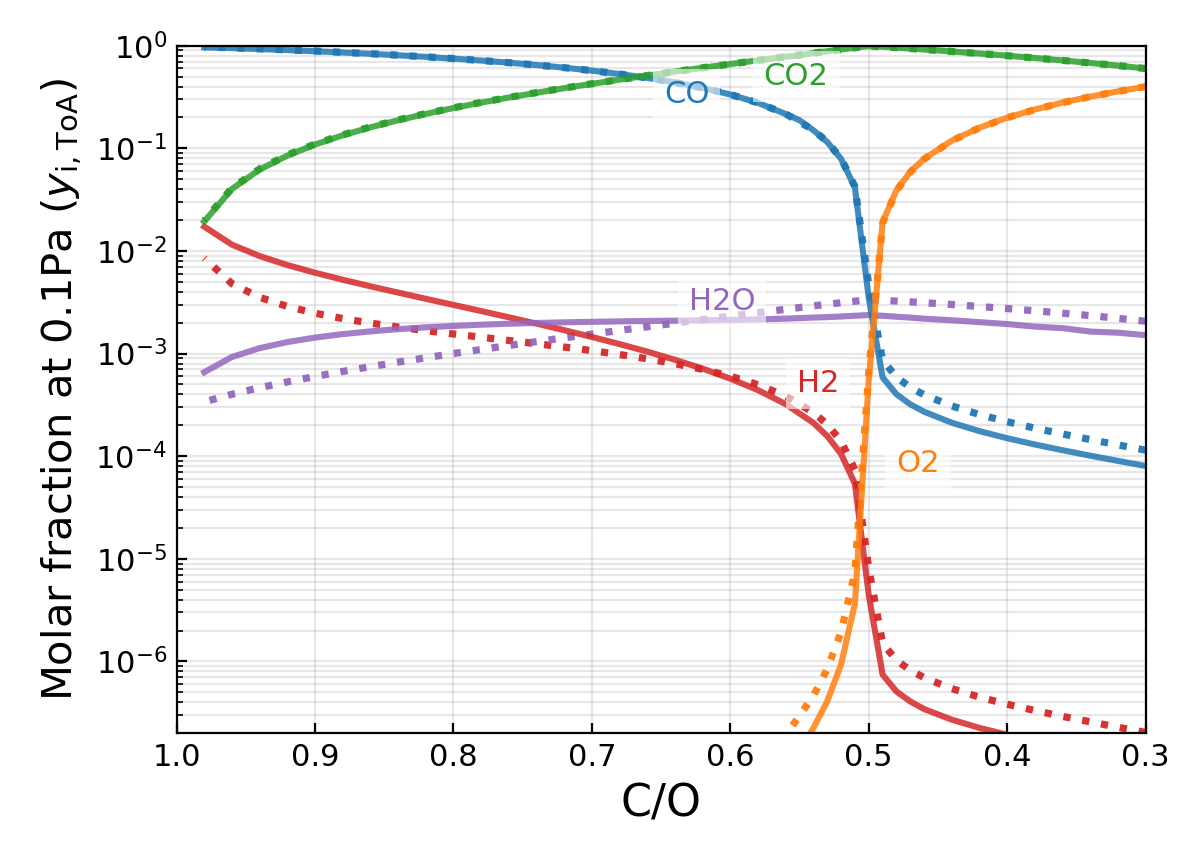}
    \caption{Effect of vertical mixing on the top-of-atmosphere mole fractions $y_{i,\rm ToA}$, evaluated at $P=0.1~\mathrm{Pa}$ for the case of a 55~Cnc~e–like planet with $T_{\rm eq}=1200\,\mathrm{K}$.
    The model parameters, colours and line styles are the same as in Fig.~\ref{figure4}.}
    \label{figure_ToA_yi}
\end{figure}

The mantle redox state of a MO exoplanet shapes the atmospheric composition of a ``puffy Venus''.
Vertical mixing further modifies the coupled atmosphere--MO structure through energy transport.
We quantify these effects using a fiducial planet motivated by 55~Cnc~e, with $M_{\rm s}=7.9\,M_{\oplus}$ and an atmosphere--MO interface radius of $R_{\rm s}=1.75\,R_{\oplus}$.
The value of $R_{\rm s}$ is calculated from Equation~(\ref{eq:Ms_Rs}) using the mass--radius prescription of \citet{2024ApJ...Peng}.
We vary C/O from $0.98$ to $0.30$, spanning moderately reducing to highly oxidizing MO conditions, and adopt $K_{\rm zz}^{\rm heat}=0$, $10^{5}$ and $10^{8}\ {\rm cm^{2}\,s^{-1}}$.
Fig.~\ref{figure4} shows how C/O and vertical mixing affect the atmospheric thermal structure, MO--atmosphere exchange, atmospheric column mass, near-surface speciation, and volatile storage in the MO.
In contrast to the near-linear decrease in $P_{\rm s}$ with increasing C/O shown for Earth-mass planets in Fig.~3a of \citet{2024ApJ...Peng}, this more massive planet does not follow the same trend.

As described in Section~\ref{apdx:dissolution}, the dissolved mass fraction $X_i$ of each volatile species depends on the interface pressure $P_{\rm s}$ and the local mole fraction $y_{i,\rm s}$. We therefore examine $P_{\rm s}$ and the species partial pressures $P_{i,\rm s}=y_{i,\rm s}P_{\rm s}$ before discussing the dissolved volatile masses. This links the gas-side pressure and composition to the melt-side volatile reservoir.

Fig.~\ref{figure4}~{\bf (A)}~and~{\bf (C)} show how vertical mixing changes the gas-side conditions at the MO surface.
Fig.~\ref{figure4}~{\bf (A)} gives the total interface pressure $P_{\rm s}$, while Fig.~\ref{figure4}~{\bf (C)} decomposes it into species partial pressures $P_{i,\rm s}$.
Without mixing, \ce{CO} provides the largest pressure contribution for C/O $\simeq 0.6$--1, \ce{CO2} contributes mainly near C/O $\simeq 0.5$, \ce{O2} dominates at the oxidizing end, and \ce{H2} is important only at the most reducing end.
Because atmospheric C/O is used here as a proxy for mantle redox state, these no-mixing speciation trends are broadly consistent with previous magma-ocean studies showing that the relative abundances of reduced and oxidized gases depend on temperature and oxygen fugacity \citep{Sossi2020SciA,Nicholls2026CHILI}.
The corresponding $P_{\rm s}$ decreases with C/O for C/O $<0.5$, but increases for C/O $>0.5$.

Vertical mixing changes both $P_{\rm s}$ and its partitioning among species. 
For $K_{\rm zz}^{\rm heat}=10^{5}\ {\rm cm^{2}\ s^{-1}}$, mixing generally raises $P_{\rm s}$, especially at low C/O. 
For $K_{\rm zz}^{\rm heat}=10^{8}\ {\rm cm^{2}\ s^{-1}}$, it further raises $P_{\rm s}$ at the oxidizing end, but can reduce it relative to the no-mixing case for C/O $>0.5$. 
Mixing also increases the near-surface partial pressures of \ce{O2} and \ce{H2}, reduces \ce{CO2}, and shifts CO in opposite directions: CO increases at the oxidizing end but decreases slightly for C/O $\simeq 0.6$--1. 
Thus, mixing does not simply increase the atmospheric column mass. 
It redistributes the gas-side pressure among volatile species and changes which species dominate dissolution into the MO.

Fig.~\ref{figure4}~{\bf (B)} shows the resulting volatile masses dissolved in the MO. 
At fixed C/O, hydrogen is preferentially retained in the MO, mainly as dissolved \ce{H2O}, whereas carbon is more strongly partitioned into the atmosphere. 
Mixing increases the dissolved masses of \ce{H2} and CO in the MO, while \ce{CO2} is preferentially retained in the atmosphere. 
This contrast is strongest at high C/O.

Fig.~\ref{figure4}~{\bf (D)} shows the surface temperature $T_{\rm s}$.
The temperature turns over near C/O $=0.5$.
For C/O $<0.5$, $T_{\rm s}$ decreases with increasing C/O, and mixing raises $T_{\rm s}$.
For C/O $>0.5$, $T_{\rm s}$ increases with C/O, and stronger mixing gives a larger temperature increase.
The increase in $T_{\rm s}$ is driven by the mechanically induced downward flux $F_{\rm mix}$.
Its pressure derivative, $dF_{\rm mix}/dP$, is the only term through which  $K_{\rm zz}^{\rm heat}$ directly enters the energy balance.
The radiative and convective fluxes adjust to the resulting thermal structure.
Here,  $K_{\rm zz}^{\rm heat}$ does not directly transport chemical species.
Instead, their abundances respond to the modified temperature through thermochemical equilibrium.
During the structural calculation, these abundance changes do not feed back through opacity because the mean opacities are fixed.
In the subsequent spectral calculation, the converged T--P and abundance profiles are used to compute wavelength-dependent opacities and emission spectra.

In the strongest-mixing cases, the near-surface temperature can reach regimes where rock vapour may become important, often at $T\gtrsim5000\,{\rm K}$.
All atmospheric layers considered here remain above $500~\mathrm{K}$.
Gas--liquid condensation is therefore not expected for \ce{H2}, \ce{O2}, \ce{CO}, \ce{CO2}, or \ce{CH4}.
\ce{H2O} condensation could in principle occur where $T<647~\mathrm{K}$ and its partial pressure exceeds the saturation vapour pressure, but this process is not included or explicitly evaluated in the present model.
Condensable rock vapour could introduce additional compositional gradients and modify the deep $P$--$T$ structure
\citep{Misener2023}.
For the hottest planets, the MO surface may approach or exceed the silicate critical region, making the predicted pressure, composition and atmospheric thickness sensitive to the equation of state, chemical network and definition of the atmosphere--magma boundary.
In our hottest models, Fig.~\ref{figure4}~{\bf (A)} and~{\bf (D)} show that
$P_{\rm s}\gtrsim10^4~\mathrm{bar}$ and $T_{\rm s}>7000~\mathrm{K}$.
These conditions exceed the estimated critical pressure and temperature ranges of
$80$--$130~\mathrm{MPa}$ and $6500$--$7000~\mathrm{K}$ for bulk-silicate-Earth-like material \citep{Caracas2023}, indicating that the nominal atmosphere--MO interface enters the estimated supercritical-silicate regime.
Further \textit{ab initio} calculations and high-pressure experiments are needed to test this regime.

In summary, vertical mixing is a key mediator between the MO redox state and the atmospheric structure of a ``puffy Venus''. By coupling transport and thermochemistry at the MO--atmosphere interface, mixing reshapes the atmospheric column mass, near-surface composition, volatile partitioning, and surface temperature in a C/O-dependent way (Fig.~\ref{figure4}). It generally raises $P_{\rm s}$ and $T_{\rm s}$, but vigorous mixing can reduce $P_{\rm s}$ for C/O $>0.5$, showing that mixing changes not only the atmospheric column mass but also the distribution of volatiles between the atmosphere and the MO.

\subsection{The effect of mixing on the emission spectrum}\label{Sec:mixing on spectrum}

\begin{figure*}
    \includegraphics[width=1.0\linewidth]{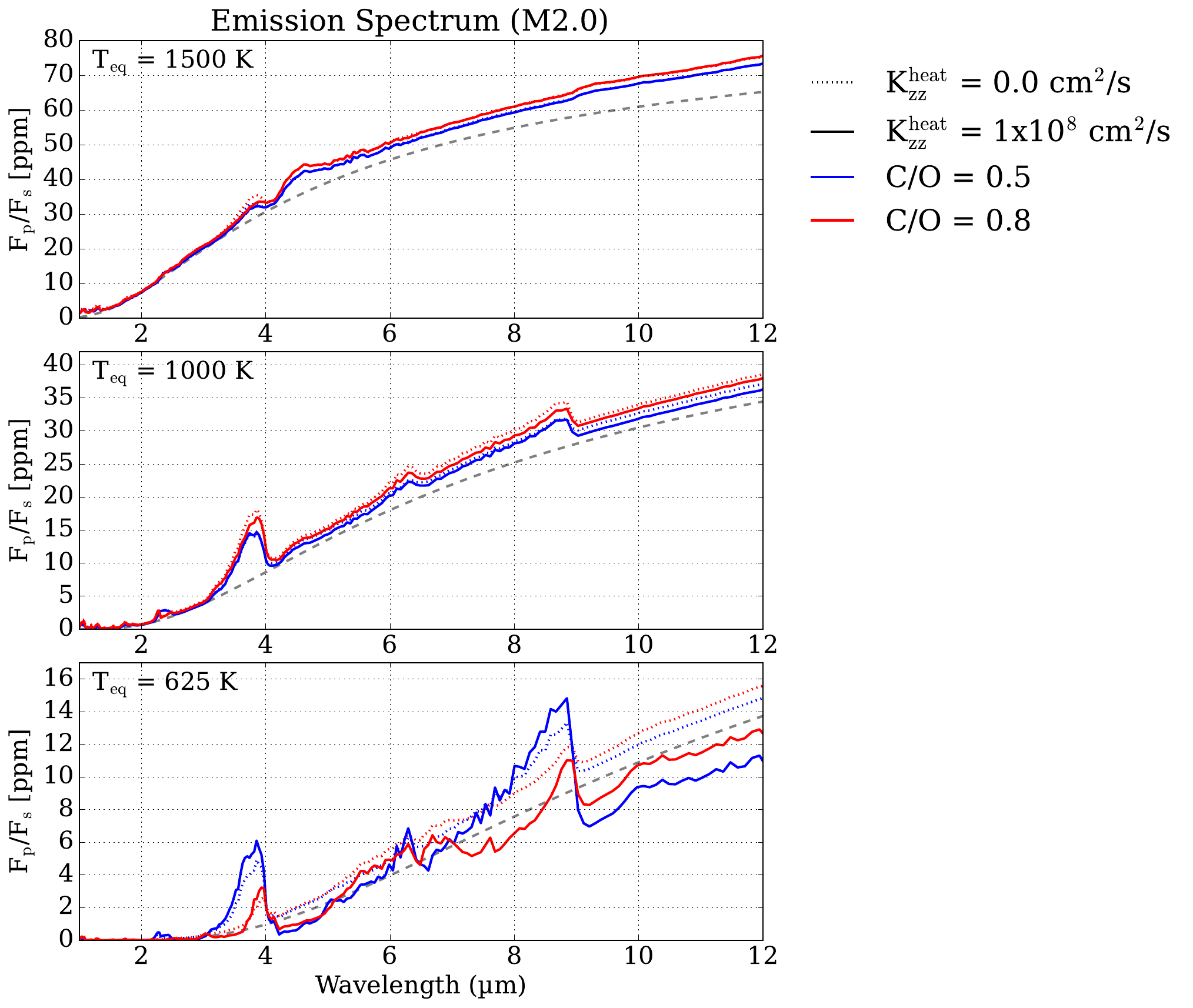}
    \caption{
    Planet--star flux ratio for the puffy-Venuse (M2.0, 2.0~$M_\oplus$) showing the effect of vertical mixing: solid and dotted lines correspond to models with and without mixing respectively, gray dashed lines indicate blackbody emission, panels from top to bottom show $T_{\rm eq} = 1500$, 1000, and 625~K, and blue/red lines represent C/O = 0.5/0.8.
    }
    \label{emission_M2}
\end{figure*}

\begin{figure*}
    \centering
    \includegraphics[width=1.0\linewidth]{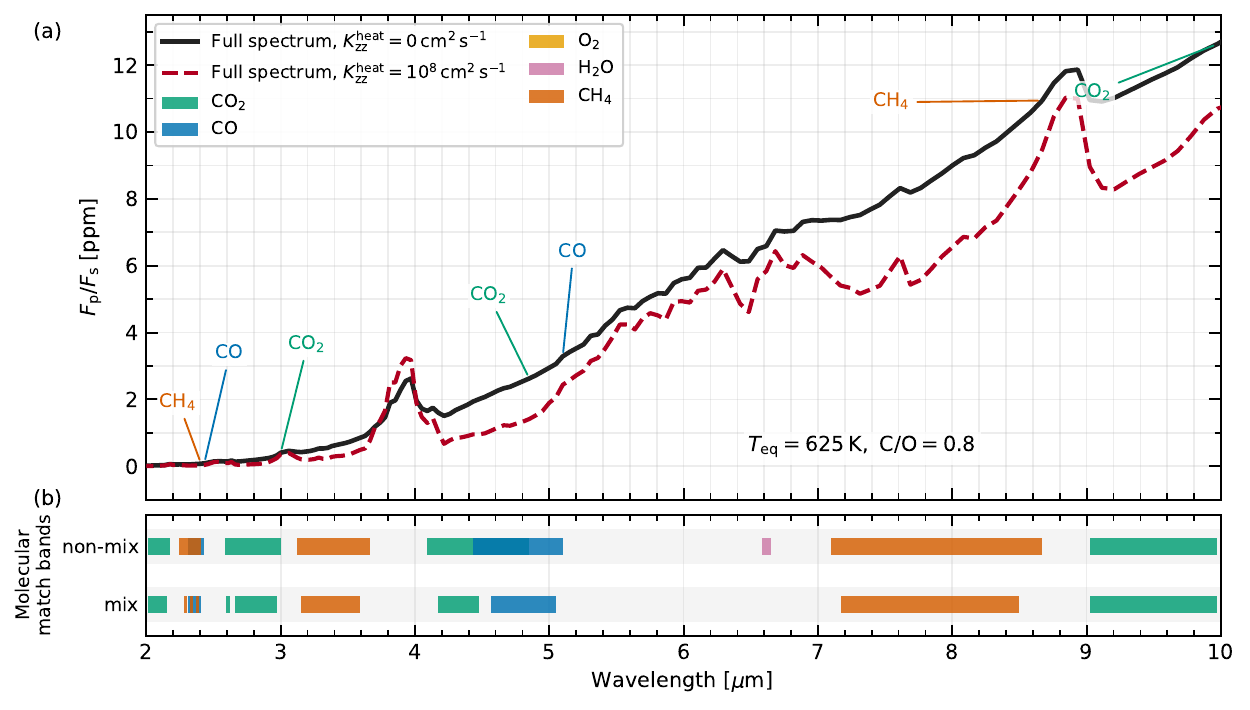}
    \caption{
    Emission spectra and opacity-source diagnostics for the M2.0 model with C/O = 0.8 and \(T_{\rm eq}=625\) K. 
                    (a) Full planet-to-star flux spectra without vertical mixing $ (K_{\rm zz}^{\rm heat}=0~{\rm cm^2~s^{-1}}$, solid black) and with vertical mixing  \( K_{\rm zz}^{\rm heat}=10^8~{\rm cm^2~s^{-1}}\), dashed red).  
                    Molecular labels mark representative wavelength regions where individual-opacity spectra reproduce features in the full spectrum. 
                    (b) Barcode-style match bands show wavelength intervals where spectra computed with only one molecular opacity source retained closely match the corresponding full spectrum for the non-mixed and mixed cases. 
                    Colors indicate \ce{CO2}, \ce{CO}, \ce{O2}, \ce{H2O}, and \ce{CH4}; gray regions indicate wavelengths where no single molecular opacity source alone closely reproduces the full spectrum, suggesting combined opacity effects from multiple sources. 
                    Compared with the case without vertical mixing, an expanded or more prominent match band in the mixed case can indirectly indicate an enhanced influence of the corresponding opacity source on the emergent spectrum. 
                    The match bands are diagnostic and should not be interpreted as direct opacity strengths or additive molecular contributions.}
    \label{fig_single_species}
\end{figure*}

\begin{figure}
    \centering
    \includegraphics[width=1.0\linewidth]{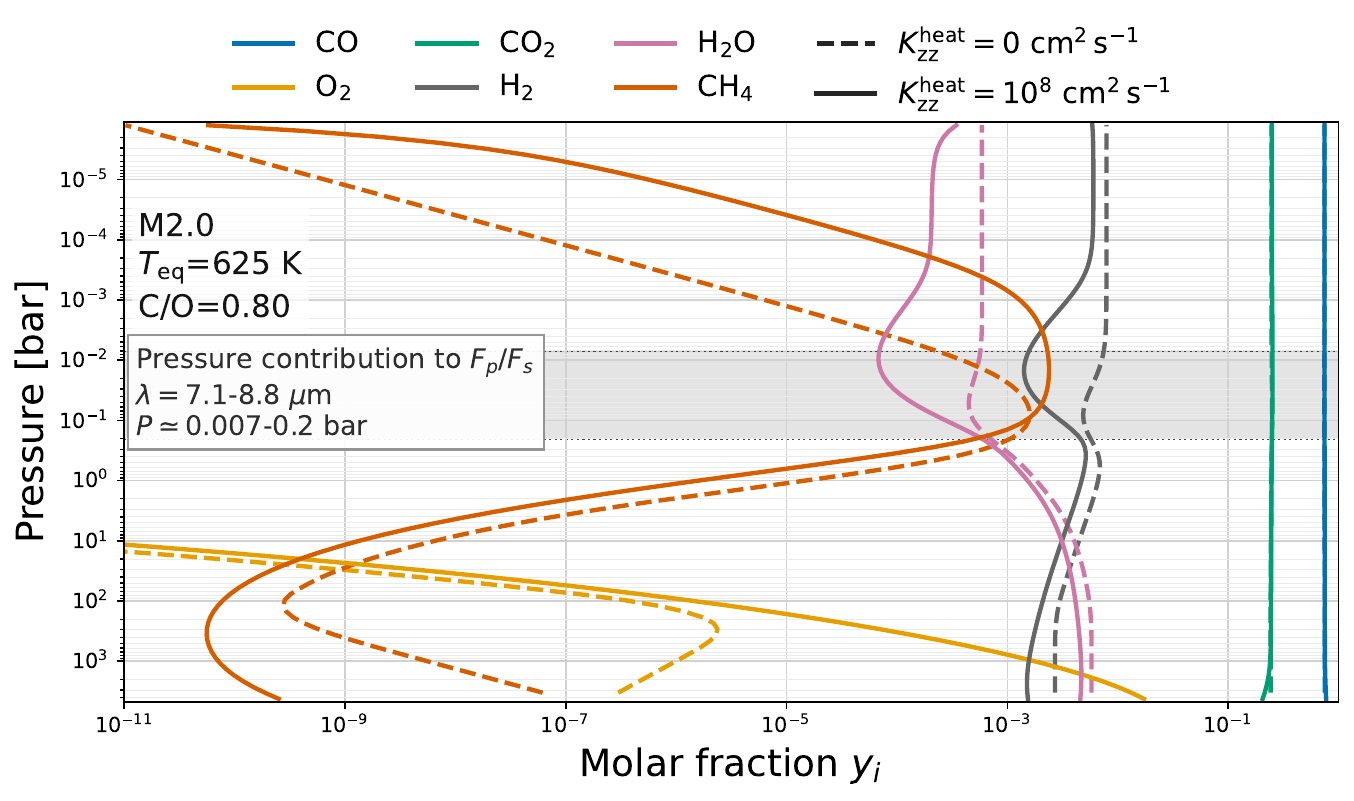}
    \caption{
            Molar fraction profiles for the $M2.0$, $T_{\rm eq}=625\,{\rm K}$, ${\rm C/O}=0.80$ case. Solid and dashed lines show $K_{\rm zz}^{\rm heat} = 10^8$ and $0\,{\rm cm^2\,s^{-1}}$, respectively. The grey region marks the pressure range ($P\simeq0.007$--$0.2\,{\rm bar}$) contributing most to $F_{\rm p}/F_{\rm s}$ at $7.1$--$8.8\,\mu{\rm m}$. Molecular colors match Fig.~\ref{fig_single_species}.
        }
    \label{Mcore2.00_yi_profile_CO0.80_Teq625}
\end{figure}

\begin{figure*}
    \includegraphics[width=1.0\linewidth]{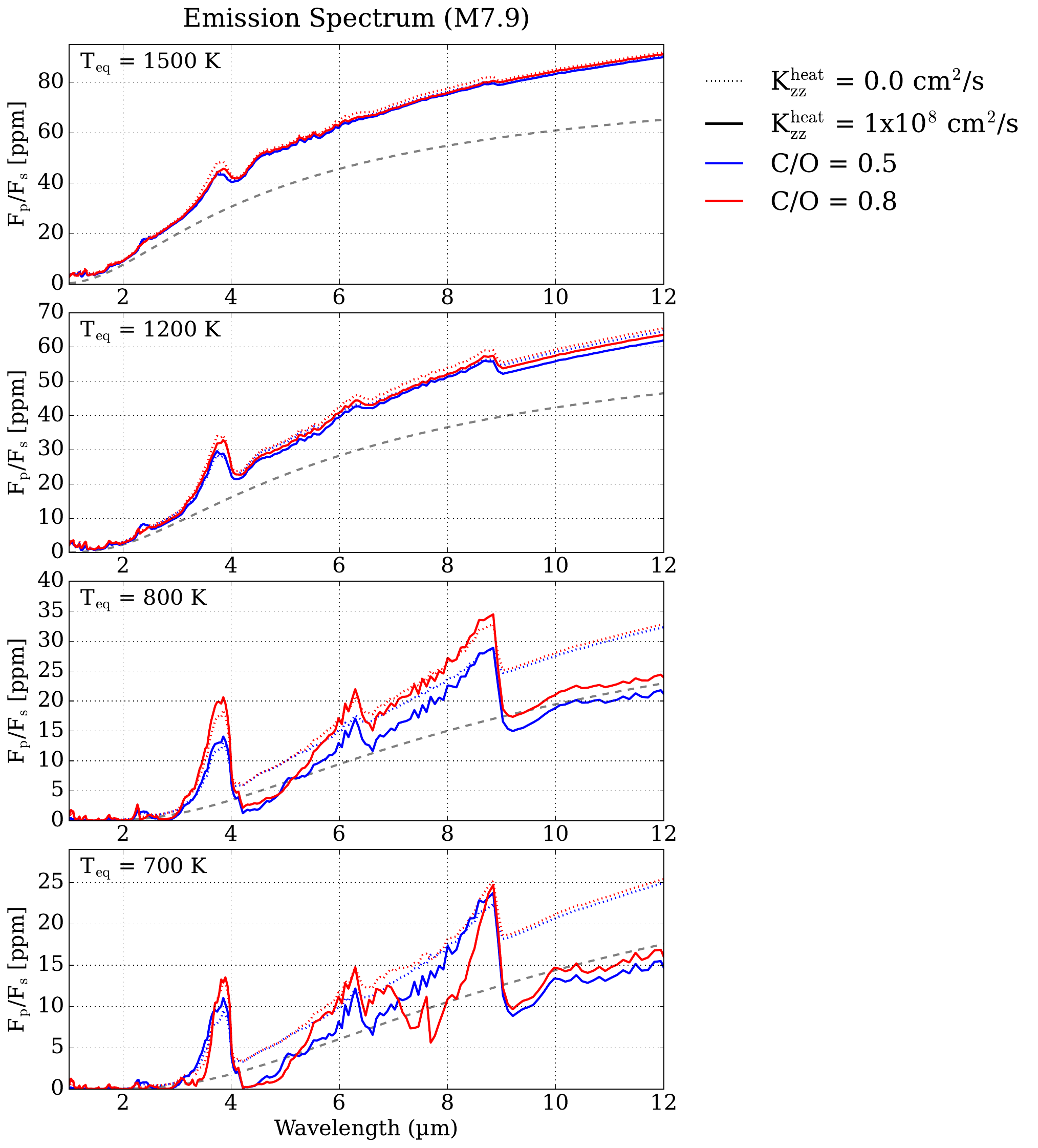}
    \caption{
    Effect of vertical mixing on the emission spectrum (planet--star flux ratio) for M7.9  ($7.9\,M_\oplus$). 
    The calculations adopt the planetary and stellar parameters of 55~Cnc~e. 
    Solid and dotted lines show models with and without vertical mixing, respectively, and the gray dashed line marks blackbody emission. 
    Panels from top to bottom correspond to equilibrium temperatures of 1500, 1200, 800, and 700~K. 
    Colors are the same as in Fig.~\ref{emission_M2}.}
    \label{figure_spectrum79}
\end{figure*}

\begin{figure*}
  \centering
  \includegraphics[width=1.0\textwidth]{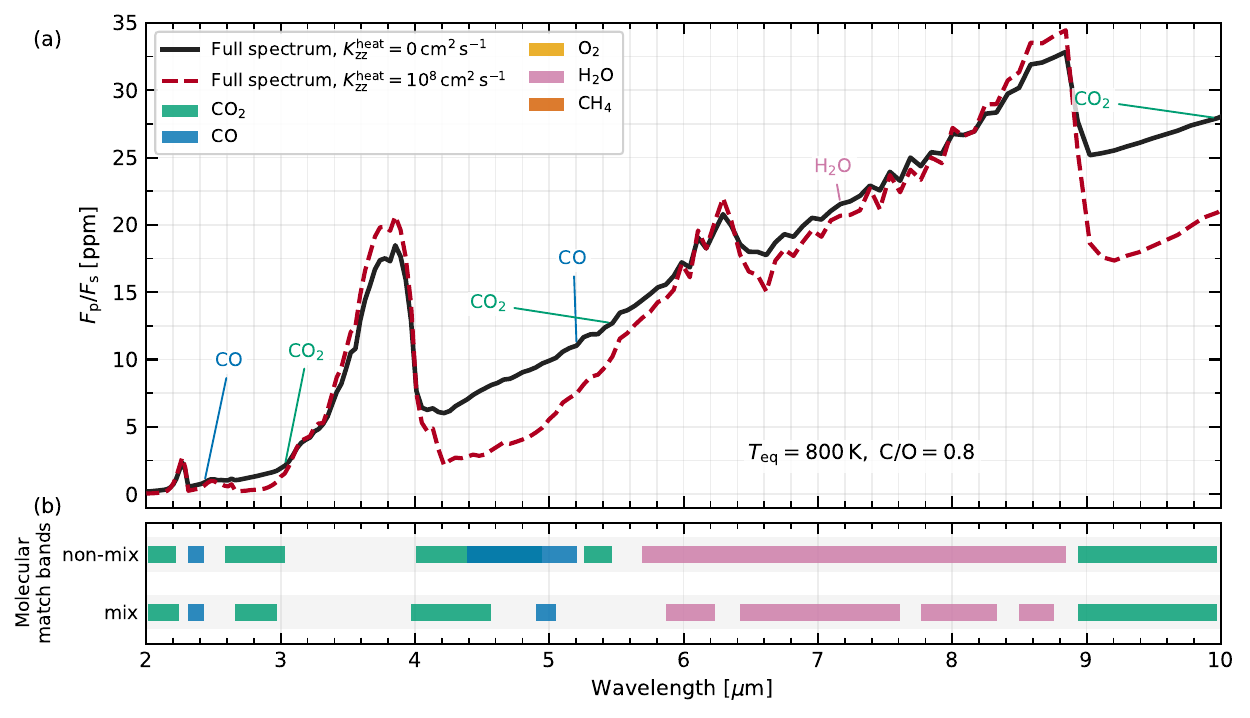}
  \caption{ Same as Fig.~\ref{fig_single_species}, but for the M7.9 model with C/O = 0.8 and $T_{\rm eq}=800,\mathrm{K}$, comparing $K_{\rm zz}^{\rm heat}0$ and $10^8\,\mathrm{cm^2\,s^{-1}}$.}
  \label{fig_M79_800K_single_species_spectra_0}
\end{figure*}

\begin{figure*}
  \centering
  \includegraphics[width=1.0\textwidth]{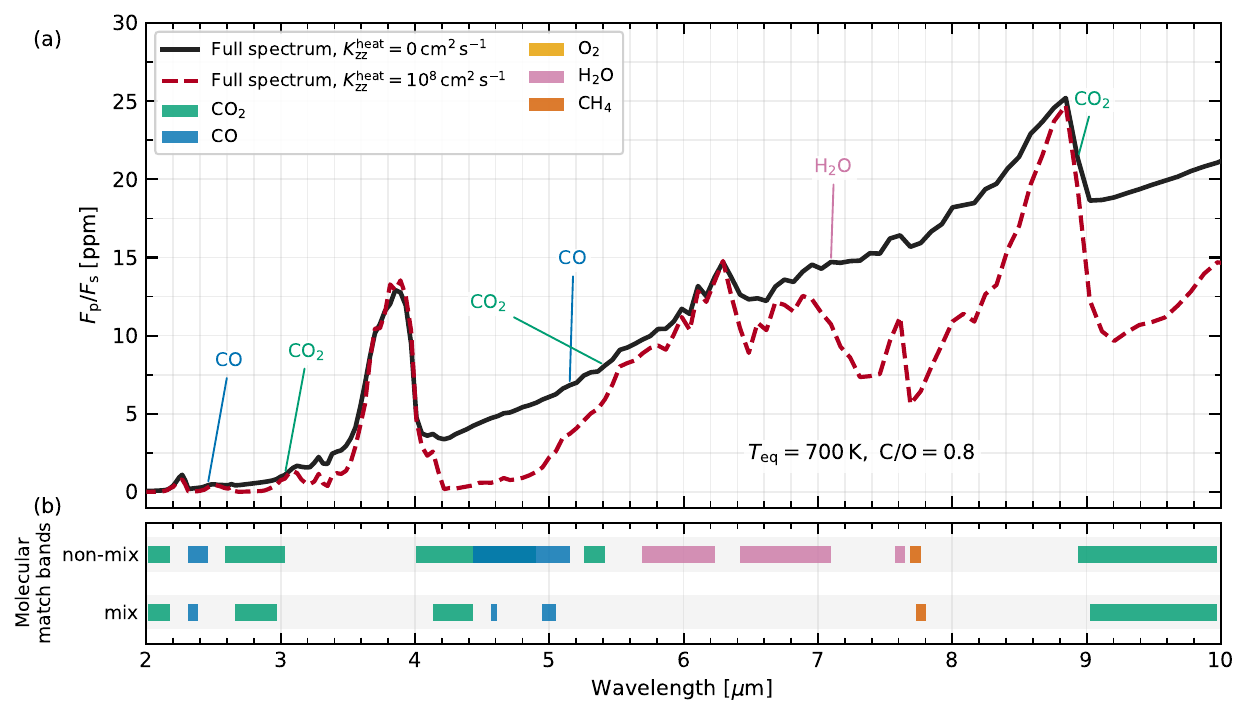}
  \caption{Same as Fig.~\ref{fig_single_species}, but for the M7.9 model with C/O = 0.8 and $T_{\rm eq}=700,\mathrm{K}$, comparing $K_{\rm zz}^{\rm heat} = 0$ and $10^8\,\mathrm{cm^2\,s^{-1}}$.}
  \label{fig_M79_700K_single_species_spectra_1}
\end{figure*}

Before turning to the emission spectra, we first examine the upper-atmosphere composition for the same model grid as in Fig.~\ref{figure4}.
We evaluate the mole fractions $y_{i,\rm ToA}$ at $P=0.1~\mathrm{Pa}$, the upper boundary adopted in our model and by \citet{2024ApJ...Peng}.
For the high-$T_{\rm eq}$ cases considered here, a comparison with $P=100~\mathrm{Pa}$ gives the same qualitative trends because the abundance changes induced by mixing-related heating are small at both pressure levels.
In the no-mixing baseline, \ce{CO} is favoured at high C/O, \ce{O2} at low C/O, \ce{CO2} peaks near C/O $\approx 0.5$, and \ce{H2} becomes appreciable mainly at the most reducing end.
With mixing, the ToA abundances of \ce{CO} and \ce{O2} change only modestly, whereas \ce{H2O} and \ce{H2} are more sensitive and show a turnover near C/O $\approx 0.6$.
Fig.~\ref{figure_ToA_yi} therefore provides a common upper-atmosphere compositional diagnostic.
The $0.1~\mathrm{Pa}$ level is not intended to represent a characteristic emission pressure, and the spectra are calculated using the full atmospheric profiles.

We use the calculated temperature-pressure profiles and elemental abundances as inputs to the 1D radiative transfer forward model \texttt{POSEIDON}\footnote{\url{https://github.com/MartianColonist/POSEIDON}} python package \citep{MacDonald_and_Madhusudhan_2017,MacDonald_2023,Coulombe_etal2023}. Emission spectra are then calculated assuming the planetary parameters of TOI-561~b and 55 Cnc~e, together with the stellar parameters of their respective host stars. We note that since the opacity database used by \texttt{POSEIDON} is limited to temperatures below 2000\,K and pressures below 100\,bar, we restrict our spectra calculations to atmospheric layers with pressures $\leq$100\,bar. This is unlikely to affect the observable spectrum, as the photosphere typically resides at pressures $\leq$100\,bar. For atmospheric regions where the temperature exceeds 2000\,K in the deeper atmosphere, we assume an isothermal profile of 2000\,K.

Figure~\ref{emission_M2} shows the thermal emission spectra of the
$2.0\,M_{\oplus}$ model (M2.0), presented as a TOI-561~b-like case study.
For this case, we adopt a planetary radius of
$R_{\rm p}=1.397\,R_{\oplus}$ and a stellar radius of
$R_{\star}=0.843\,R_{\odot}$ \citep{Piotto2024}.
The host star has an effective temperature of
$T_{\rm eff,\star}=5432~\mathrm{K}$, a metallicity of
$\mathrm{[Fe/H]}=-0.303$, and a surface gravity of
$\log g_{\star}=4.451$ (cgs) \citep{MacDougall2023}.
The spectra are plotted as the planet-to-star flux ratio, $F_{\rm p}/F_\star$, as a function of wavelength.
In the high-$T_{\rm eq}$ case ($T_{\rm eq}=1500\,\mathrm{K}$), vertical mixing has little effect on the spectrum: for a fixed C/O ratio, the mixed and non-mixed cases are nearly identical.
This suggests that, at high $T_{\rm eq}$, the mixing-induced changes in thermal structure and equilibrium chemistry are too small to produce an appreciable spectral response.
At the intermediate equilibrium temperature, $T_{\rm eq}=1200\,\mathrm{K}$, vertical mixing has a stronger spectral effect.
The mixed cases show a lower $F_{\rm p}/F_\star$ than the non-mixed cases for all adopted C/O ratios.
In the low-$T_{\rm eq}$ case ($T_{\rm eq}=625\,\mathrm{K}$), vertical mixing produces a strong and wavelength-dependent change in the emission spectrum.
For C/O = 0.5, mixing increases $F_{\rm p}/F_\star$ near $3.5$--$4.0\,\mu\mathrm{m}$ and $7.5$--$9\,\mu\mathrm{m}$, indicating higher atmospheric transparency and lower effective opacity in these bands.
At other wavelengths, especially between $9$ and $12\,\mu\mathrm{m}$, the flux decreases, indicating lower transparency and higher effective opacity.
For C/O = 0.8, the effect is stronger: mixing suppresses $F_{\rm p}/F_\star$ across most of $4$--$12\,\mu\mathrm{m}$.

This increasing sensitivity to vertical mixing toward lower $T_{\rm eq}$ reflects the temperature dependence of the opacity. 
In the high-$T_{\rm eq}$ model, the opacity structure depends only weakly on $T_{\rm eq}$, consistent with the electron-scattering opacity becoming saturated. 
The abundance changes introduced by vertical mixing therefore do not substantially modify the wavelength-dependent effective opacity in the mid-infrared emitting regions, and the mixed and non-mixed spectra remain similar. 
At lower $T_{\rm eq}$, the opacity varies more steeply with the local thermal and chemical structure. 
Vertical mixing changes the mole fractions of the main molecular absorbers within the pressure ranges sampled by the emission spectrum, producing larger variations in effective opacity, reducing atmospheric transparency and suppressing $F_{\rm p}/F_\star$. 
Because this suppression is strongest for C/O = 0.8 in the low-$T_{\rm eq}$ case, we focus on this case below.

The lower mid-infrared flux in the mixed C/O = 0.8 spectrum is therefore consistent with the larger opacity changes discussed above.
However, the modified $T$--$P$ structure can also affect the emergent spectrum through the thermal source function.
Because the opacity and thermal-structure effects are coupled in the present calculations, we do not quantify their relative contributions here.
The following single-species spectra are therefore used to identify the molecular opacity sources associated with the wavelength-dependent spectral features, rather than to separate the opacity contribution from the effect of the modified thermal structure.
The resulting equilibrium abundance changes
do not create new spectral bands, but change the relative contributions of existing bands from \ce{CO2}, \ce{CO}, and \ce{CH4}.
Where the effective opacity increases, the atmosphere becomes less transparent and the emergent flux decreases.
Where the opacity is reduced, radiation can emerge from deeper and hotter layers, producing a local flux increase.

Vertical mixing does not create new spectral bands, but changes the relative contributions of existing bands from \ce{CO2}, \ce{CO}, and \ce{CH4}.
To identify the molecular origin of the opacity variations, we examine the single-species spectra in Fig.~\ref{fig_single_species}.
This diagnostic is not a linear decomposition of the full spectrum (Fig.~\ref{fig_single_species}a).
Instead, it identifies the dominant molecular contributors to the opacity over different wavelength intervals.
The primary differences occur in the carbon-bearing molecules shown in Fig.~\ref{fig_single_species}b.
\ce{CO2} and \ce{CO} dominate much of the $4.1$--$5.2\,\mu\mathrm{m}$ region, whereas \ce{CH4} becomes important near $3.1$--$4.0\,\mu\mathrm{m}$ and $7.1$--$8.8\,\mu\mathrm{m}$.
\ce{CO2} remains important at $9$--$10\,\mu\mathrm{m}$.

In the M2.0 model with C/O = 0.8, vertical mixing produces its strongest local suppression of $F_{\rm p}/F_\star$ in the $7.1$--$8.8\,\mu\mathrm{m}$ band. 
This is the same wavelength interval in which the single-species diagnostic identifies \ce{CH4} as an important opacity source. 
To explain why this \ce{CH4} feature appears in the low-$T_{\rm eq}$ spectrum but is not prominent in the high-$T_{\rm eq}$ spectra, we examine the vertical mole-fraction profiles in Fig.~\ref{Mcore2.00_yi_profile_CO0.80_Teq625}. 
The full $y_i$--$P$ profiles are shown because the emission at a given wavelength is shaped by a finite pressure range, rather than by the abundance at a single ToA level. 
The grey band in Fig.~\ref{Mcore2.00_yi_profile_CO0.80_Teq625} marks the pressure region that contributes most strongly to the $7.1$--$8.8\,\mu\mathrm{m}$ emission, approximately $0.007$--$0.2\,\mathrm{bar}$. 
Within this region, the $K_{\rm zz}^{\rm heat}=10^8\,\mathrm{cm^2\,s^{-1}}$ model has a higher \ce{CH4} mole fraction than the no-mixing case. 
The enhanced \ce{CH4} abundance increases the effective opacity in the same pressure region sampled by this wavelength band, explaining why vertical mixing suppresses $F_{\rm p}/F_\star$ most strongly at $7.1$--$8.8\,\mu\mathrm{m}$. 
The \ce{CH4} feature in Fig.~\ref{fig_single_species} is therefore not abrupt, it reflects the low-$T_{\rm eq}$ enhancement of \ce{CH4} opacity within the relevant emitting layers.

With mixing, the single-species matches become narrower or less continuous in several wavelength intervals. 
This does not imply weaker total opacity. 
Instead, it shows that the opacity is no longer controlled by one molecule over a broad wavelength range. 
The mixed full spectrum remains fainter because the combined effective opacity of overlapping \ce{CO2}, \ce{CO} and \ce{CH4} bands is stronger.
Vertical mixing therefore redistributes the opacity contribution among carbon-bearing molecules and suppresses the C/O = 0.8 mid-infrared emission.

To test whether the same mechanism operates at higher planetary mass, we calculated the emission spectra for the $7.9\,M_{\oplus}$ model (M7.9; Fig.~\ref{figure_spectrum79}), representing a 55 Cnc~e-like planet. 
For this case, we adopt a planetary radius of
$R_{\rm p}=1.875\,R_{\oplus}$ \citep{Bourrier2018}.
The host-star parameters are
$R_{\star}=0.980\,R_{\odot}$,
$T_{\rm eff,\star}=5198~\mathrm{K}$, and
$\log g_{\star}=4.30$ (cgs) \citep{Moutou2026},
with a metallicity of $\mathrm{[Fe/H]}=0.37$ \citep{Zhao2023}.
The overall spectral behavior is similar to that of M2.0, but the response to vertical mixing is larger. 
This stronger response reflects the combined effects of the larger gravity and the different internal structure of the M7.9 planet.
For the same  $K_{\rm zz}^{\rm heat}$, the larger gravity directly enhances the mixing-induced heat transport through Equation~(\ref{eq:Fmix}).
The resulting thermal perturbation, acting on the different hydrostatic structure of the more massive planet, leads to larger changes in the C--H--O equilibrium abundances and molecular opacity.

At $T_{\rm eq}=1500\,\mathrm{K}$, the M7.9 mixed and non-mixed spectra remain nearly identical.
As in M2.0, this indicates that vertical mixing does not strongly modify the wavelength-dependent effective opacity when $T_{\rm eq}$ is high.
As $T_{\rm eq}$ decreases, the effect of vertical mixing becomes progressively stronger, producing increasingly large spectral differences. 
The M7.9 sequence therefore follows the same high--low $T_{\rm eq}$ contrast found in M2.0. 
At high $T_{\rm eq}$, the opacity is only weakly affected by the abundance changes introduced by vertical mixing. 
At lower $T_{\rm eq}$, the opacity responds more strongly to the local thermal and chemical structure, so mixing-induced changes in molecular abundances produce larger changes in atmospheric transparency and thermal emission.
The largest mixed--non-mixed separations occur at $T_{\rm eq}=800\,\mathrm{K}$ and $T_{\rm eq}=700\,\mathrm{K}$, particularly for C/O = 0.8. 
In these models, vertical mixing suppresses $F_{\rm p}/F_\star$ over broad mid-infrared wavelength ranges, indicating substantial changes in the atmospheric opacity structure. 
We therefore focus on the C/O = 0.8 cases to investigate the molecular origin of these spectral variations.

At $T_{\rm eq}=800,\mathrm{K}$, the M7.9 model shows \ce{CO2}, \ce{CO}, and \ce{H2O} as the main opacity sources (Fig.~\ref{fig_M79_800K_single_species_spectra_0}).
Compared with the non-mixed model, the wavelength intervals associated with these molecules become less continuous after vertical mixing, particularly for \ce{H2O}.
At the same time, the full emission spectrum exhibits both local flux enhancements and suppression.
This behavior suggests that the opacity changes associated with vertical mixing are not uniform across wavelength.
Instead, vertical mixing changes the relative importance of different opacity sources, contributing to the strongly wavelength-dependent spectral response.

At $T_{\rm eq}=700\,\mathrm{K}$, the M7.9 model shows a substantially stronger response to vertical mixing(Fig.~\ref{fig_M79_700K_single_species_spectra_1}). 
The mixed spectrum is systematically fainter across much of the $4$--$10\,\mu\mathrm{m}$ range, indicating lower atmospheric transparency and a stronger net opacity effect. 
The dominant opacity sources are primarily \ce{CO2} and \ce{CO}, while \ce{CH4} contributes over several narrower wavelength intervals. 
Compared with the $T_{\rm eq}=800\,\mathrm{K}$ model, many molecular matching regions become shorter or fragmented after mixing. 
However, the overall suppression of $F_{\rm p}/F_\star$ is significantly larger, implying that the combined opacity of multiple absorbers becomes more effective at reducing the escaping thermal radiation.

Overall, the M7.9 results support the same physical picture inferred from M2.0. 
Vertical mixing modifies the vertical abundance distributions of the major opacity carriers and changes their relative contributions to the wavelength-dependent effective opacity. 
As $T_{\rm eq}$ decreases, the opacity becomes more sensitive to the local thermal and chemical structure, so the abundance changes introduced by vertical mixing produce larger changes in atmospheric transparency and thermal emission.
The resulting spectral variations therefore arise from the combined effects of multiple overlapping opacity sources, rather than from the strengthening of a single molecular absorber alone.

Although the M2.0 and M7.9 models are motivated by TOI-561~b and 55~Cnc~e, respectively, they are not intended as best-fit models.
Our 1D calculations illustrate how gravity-wave-driven vertical mixing and its associated energy transport modify atmospheric structure, equilibrium composition, and thermal emission.
Whether the resulting spectral differences are detectable with JWST depends on the observing mode and the precision achieved.
We have not assessed this detectability and therefore do not claim that a particular JWST instrument can distinguish these differences.
Beyond our 1D treatment, future studies combining multidimensional models with phase-curve observations could investigate how atmospheric circulation shapes the spatial distribution of temperature and composition.

\section{Conclusion and discussion}\label{Sec:conclusion}

Our results show that the atmosphere--interior coupling of puffy Venuses cannot be understood from radiative transfer and chemical equilibrium alone. 
Vertical energy transport provides an additional non-radiative heating pathway, with atmospheric circulation and gravity-wave breaking representing possible dynamical origins in the radiative zone. 
This process warms the deep atmosphere and the magma-ocean surface, and lowers the stellar irradiation threshold required to maintain surface melting.
At the same time, {the associated thermal changes modify the chemical structure} of C--H--O atmospheres by enhancing \ce{H2}, \ce{O2}, and \ce{CO} in the deep atmosphere while suppressing \ce{CO2}, and by changing volatile partitioning between the atmosphere and the magma ocean.
The strength of these responses further depends on the atmospheric C/O ratio, pointing to a nonlinear coupling among atmospheric dynamics, volatile speciation, and planetary redox state.

These findings yield observable signatures. 
Vertical energy transport ($K_{\rm zz}^{\rm heat}$) strongly alters the thermal emission spectra of highly irradiated rocky planets by changing key species abundances and redistributing atmospheric heat. 
This effect is strongest in the higher-mass $7.9\,M_\oplus$ model at low temperature, around $T_{\rm eq}\approx700\,\mathrm{K}$. 
In this regime, the associated thermal changes strongly modify the equilibrium distribution of the main C--H--O species.
For high C/O, these changes reduce mid-infrared transparency by redistributing the opacity contributions of \ce{CO}, \ce{CO2}, and \ce{H2O}, thereby suppressing $F_{\rm p}/F_\star$ over a broad wavelength range. 
Thus, vertical energy transport has a stronger impact on the thermal emission spectrum than in the $2.0\,M_\oplus$ model.

The vertical energy transport considered here is therefore not a minor correction in lava-world atmosphere models.
Rather, it can substantially modify the vertical temperature structure, equilibrium atmospheric composition, and mid- to far-infrared emission spectra of these planets.
These mass-dependent differences suggest that vertical energy transport can produce more prominent spectral signatures in higher-mass planets such as 55 Cnc~e compared to lower-mass worlds like TOI-561~b.
Such differences may provide useful targets for future infrared observations, although their detectability will depend on observational precision and instrumental characteristics.

This study nonetheless has several limitations. The current framework does not yet include a multi-band one-dimensional spectral treatment \citep{2017MNRAS...Gandhi}, photochemistry \citep{2026ApJ...Mang}, or chemical disequilibrium. In real atmospheres, dynamical processes can drive vertical tracer transport and thereby modify the distribution of chemical species \citep{Zhang2018Global_I,Zhang2018Global_II,Zhang_2020}. Large-scale atmospheric circulation can contribute to this transport \citep{Zhang2018Global_I,Zhang2018Global_II,Zhang_2020}, while gravity-wave breaking and wave-driven circulation may provide additional sources of vertical mixing \citep{Moses2014Chemical,Zhang_2020}. If such transport becomes faster than the chemical reactions required to maintain local equilibrium, species can become chemically quenched and depart from thermochemical equilibrium \citep{Visscher_Moses2011Quenching,Moses2014Chemical}. These compositional transport processes are not included in the present framework, in which $K_{zz}^{\rm heat}$ represents thermal-energy transport only. In our latest work, we have begun to quantify the effect of vertical mixing in one-dimensional spectral models\footnote{Zhang et al. 2026, in preparation.}. Future studies will extend the present framework by coupling it to magma ocean--mantle evolution and by including chemical disequilibrium and photochemical processes. These additions should enable a more realistic description of transmission spectra, transit radii, and the signatures of primary transit and secondary eclipse.

Among the known candidate puffed-up rocky planets, TOI-561~b and 55~Cnc~e are particularly important because both are highly irradiated and underdense. TOI-561~b supports the theoretical plausibility of radius inflation by a carbon-rich atmosphere, whereas 55~Cnc~e provides a stronger observational benchmark because JWST observations have already suggested a CO/C\ce{O2}-rich atmosphere \citep{2024ApJ...Peng}. We therefore adopt 55~Cnc~e as the main target for future work. In a coupled atmosphere--magma-ocean--mantle framework, mineral clouds may also form above the magma ocean \citep{2025ApJ...Lee} when refractory species outgas from the molten surface and condense at higher, cooler altitudes. Such clouds could further modify both the atmospheric structure and the observable spectra.

\begin{acknowledgments}
     This work has been supported by the the National Natural Science Foundation of China (Grant No. 12503068, 42578010, and T2495234), the Basic Research Program of Jiangsu (Grant No. BK20251249), the Guangdong Basic and Applied Basic Research Foundation (Grant 2023A1515110805), the Fundamental Research Funds for the Central Universities (Grant No. 2026300367), and the Science and Technology Development Fund, Macau SAR (File No. 0139/2024/RIA2). 
     C.Y. has been supported by the National SKA Program of China (Grant No. 2022SKA0120101), the National Natural Science Foundation of China (Grant No. 12288102, 11873103, and 12373071), the science research grants from the China Manned Space Project (No. CMS-CSST-2021-B09, CMS-CSST-2021-B12, and CMS-CSST-2021-A10), the International Centre of Supernovae, Yunnan Key Laboratory (No.202302AN360001). MTM acknowledges support from the Croucher Postdoctoral Fellowship, funded by the Croucher Foundation.

\end{acknowledgments}

\appendix

\section{Temperature Dependence of the Equilibrium Constants}
\label{app:equilibrium_constants}

\begin{figure}[t!]
    \centering
    \includegraphics[width=1.0\linewidth]{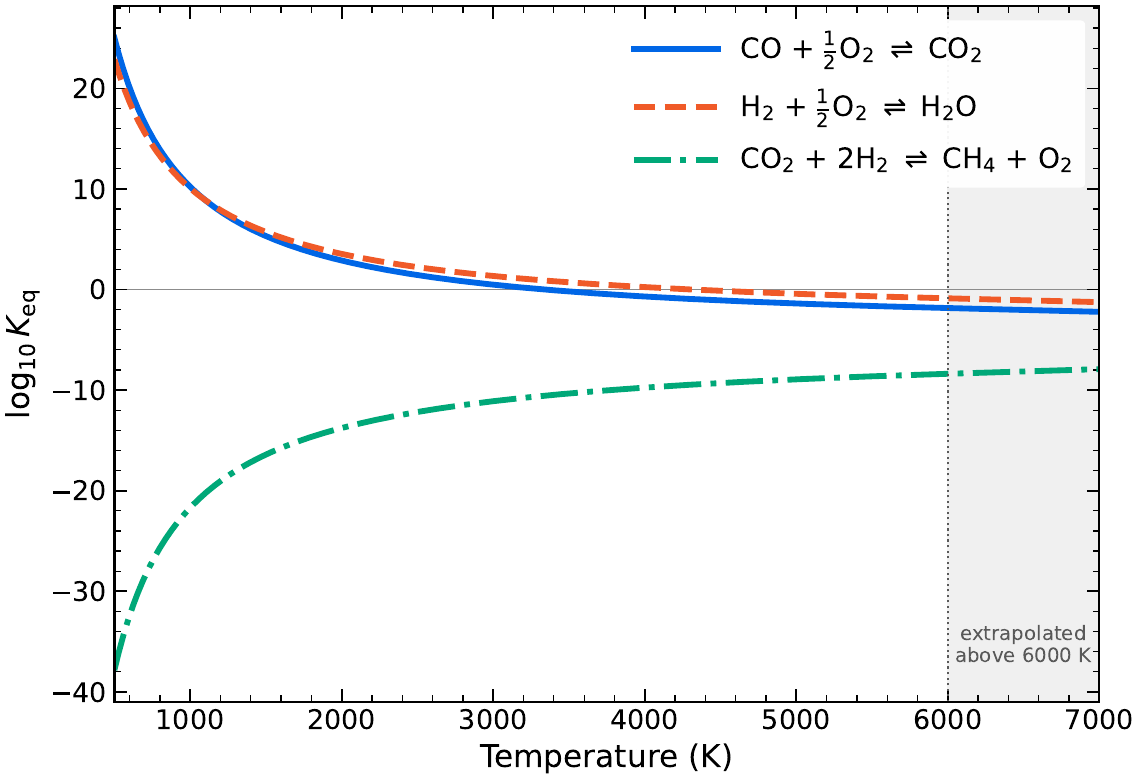}
    \caption{
    Equilibrium constants for the three gas-phase reactions as functions of temperature.
    The values are calculated from the NIST Shomate thermochemical data for a standard state of $1~\mathrm{bar}$.
    The shaded region denotes $T>6000~\mathrm{K}$, where the NIST-based thermochemical data are linearly extrapolated.
    }
    \label{fig:Keq}
\end{figure}

The equilibrium constants for \ce{CO + 0.5O2 <=> CO2} and
\ce{H2 + 0.5O2 <=> H2O} decrease with increasing temperature, whereas that for
\ce{CO2 + 2H2 <=> CH4 + O2} increases with temperature.
These curves provide a direct check of the thermochemical functions implemented in the model.
The atmospheric abundances are not determined by $K_{\rm eq}(T)$ alone, but also depend on the local pressure, C/O and C/H ratios, and non-ideal fugacities.

\bibliography{sample701}{}
\bibliographystyle{aasjournalv7}
\end{document}